\documentclass[10pt,conference, final]{IEEEtran}

\IEEEoverridecommandlockouts

\IEEEpubid{\begin{minipage}{.81\textwidth}\ \\[30pt] \centering
    \copyright~2019~IEEE. Personal use of this material is permitted. Permission from IEEE must be
    obtained for all other uses, in any current or future media, including reprinting/republishing
    this material for advertising or promotional purposes, creating new collective works, for resale
    or redistribution to servers or lists, or reuse of any copyrighted component of this work
    in other works.
  \end{minipage}}

\makeatletter
\newcommand{\thetitle}{\@maketitle}
\makeatother

\usepackage[utf8]{inputenc}
\usepackage[T1]{fontenc}
\usepackage[english]{babel}
\usepackage{microtype}
\usepackage{xargs}
\usepackage{relsize}
\usepackage{colortbl}
\usepackage[bookmarks=false, hidelinks]{hyperref}
\usepackage{multirow}
\usepackage{tabularx}
\usepackage{booktabs}
\usepackage{adjustbox}
\usepackage[colorinlistoftodos,prependcaption,textsize=tiny]{todonotes}
\newcommandx{\rbtodo}[2][1=]{\todo[linecolor=white,backgroundcolor=green!75!black,bordercolor=black,#1]{{\bf @rb:} #2}}
\newcommandx{\unsure}[2][1=]{\todo[linecolor=red,backgroundcolor=red!25,bordercolor=red,#1]{#2}}
\newcommandx{\change}[2][1=]{\todo[linecolor=blue,backgroundcolor=blue!25,bordercolor=blue,#1]{#2}}
\newcommandx{\info}[2][1=]{\todo[linecolor=gray,backgroundcolor=gray!25,bordercolor=gray,#1]{#2}}
\newcommandx{\improvement}[2][1=]{\todo[linecolor=Plum,backgroundcolor=Plum!25,bordercolor=Plum,#1]{#2}}
\newcommandx{\done}[2][1=]{\todo[linecolor=green,backgroundcolor=green!15,bordercolor=green,#1]{#2}}

\usepackage[most]{tcolorbox}

\usepackage{caption}
\usepackage{subcaption}
\usepackage{listing}
\newcommand{\cinl}[1]{\lstinline[language=c, basicstyle=\footnotesize]$#1$}

\usepackage{amsmath}
\usepackage{amssymb}
\usepackage{mathpartir}
\usepackage{bm}

\usepackage[inline]{enumitem}

\def\approach/{{\scshape TInA}}

\usepackage[capitalize, nameinlink]{cleveref}
\crefname{section}{Sec.}{Sec.}
\crefname{algorithm}{Alg.}{Alg.}
\crefname{figure}{Fig.}{Fig.}
\crefname{proposition}{Prop.}{Prop.}
\crefname{table}{Table}{Tables}
\crefname{definition}{Def.}{Def.}
\crefname{theorem}{Thm.}{Thm.}

\renewcommand{\paragraph}[1]{\medskip \noindent {\bf #1.}}

\newcommandx{\step}[1]{\overset{#1}{\rightarrow}}
\newcommandx{\eqv}[1][1=]{\overset{#1}{\equiv}}
\newcommandx{\arith}[1]{\text{#1}_{n}}
\newcommandx{\pointer}[1]{\text{#1}*}
\newcommandx{\struct}[1]{\text{#1}}
\newcommandx{\ext}[2]{\text{ext}_{\text{#1}, #2}\ }
\newcommandx{\restrict}[3]{#1_{#2..#3}}
\newcommandx{\bv}[1]{\overrightarrow{b_{#1}}}

\newcommand{\Rplus}{\protect\hspace{-.1em}\protect\raisebox{.35ex}{\smaller{\smaller\textbf{+}}}}

\def\Cpp/{\mbox{C\Rplus\Rplus}}
\def\na/{{\sc N/a}}

\usepackage{tikz}
\usetikzlibrary{shapes, automata, arrows, intersections, fit, calc, positioning}
\definecolor{darkgreen}{rgb}{0, 0.5, 0}
\definecolor{lightgreen}{rgb}{0.9,1,0.9}
\definecolor{lightred}{rgb}{1,0.7,0.7}
\definecolor{darkblue}{rgb}{0.3,0.3,1}
\definecolor{lightyellow}{rgb}{1,1,0.5}
\definecolor{darkyellow}{rgb}{0.8,0.8,0}
\definecolor{darkred}{rgb}{0.72,0.04,0.04}
\definecolor{BlueList}{HTML}{0078B4}
\definecolor{bggrey}{rgb}{0.8,0.85,0.8}
\definecolor{lightgrey}{rgb}{0.7,0.7,0.7}
\definecolor{darkgrey}{rgb}{0.4,0.4,0.4}
\definecolor{highlight}{HTML}{b00040}

\def\asm/{ASM}
\def\gcc/{{\sf GCC}}
\def\clang/{{\sf clang}}
\def\vs/{{\sf Visual Studio}}
\def\corcpp/{C/\Cpp/}
\def\binsec/{\textsc{Binsec}}

\def\arm/{{\sf ARM}}
\def\x86/{{\sf x86}}

\def\alsa/{{\sf ALSA}}
\def\ffmpeg/{{\sf ffmpeg}}
\def\gmp/{{\sf GMP}}
\def\libyuv/{{\sf libyuv}}
\def\libgcrypt/{{\sf libgcrypt}}
\def\linux-tools/{{\sf linux-tools}}
\def\udpcast/{{\sf UDPCast}}

\def\simd/{SIMD}

\usepackage{fontawesome}

\DeclareFontFamily{U}{fontawesome1}{}
\DeclareFontShape{U}{fontawesome1}{m}{n}{<->FontAwesome--fontawesomeone}{}
\DeclareFontFamily{U}{fontawesome2}{}
\DeclareFontShape{U}{fontawesome2}{m}{n}{<->FontAwesome--fontawesometwo}{}
\DeclareFontFamily{U}{fontawesome3}{}
\DeclareFontShape{U}{fontawesome3}{m}{n}{<->FontAwesome--fontawesomethree}{}

\def\good{{\color{darkgreen} \faCheck}}
\def\bad{{\color{darkred} \faClose}}

\usepackage{environ}
\newtoggle{islongpaper}

\NewEnviron{long-paper}{\iftoggle{islongpaper}{\BODY}{}}

\begin{document}
\bstctlcite{BSTcontrol}

\def\mytitle{Get rid of inline assembly through verification-oriented lifting}
\title{\mytitle}

\author{
  \IEEEauthorblockN{
    Frédéric Recoules\IEEEauthorrefmark{1},
    Sébastien Bardin\IEEEauthorrefmark{1},
    Richard Bonichon\IEEEauthorrefmark{1},
    Laurent Mounier\IEEEauthorrefmark{2} and
    Marie-Laure Potet\IEEEauthorrefmark{2}}
  \IEEEauthorblockA{
    \IEEEauthorrefmark{1}CEA LIST,
     Paris-Saclay, France \\
    firstname.lastname@cea.fr}
  \IEEEauthorblockA{
    \IEEEauthorrefmark{2}Univ. Grenoble Alpes. VERIMAG, Grenoble, France \\
    firstname.lastname@univ-grenoble-alpes.fr}
}

\maketitle

\begin{abstract}
  Formal methods for software development have made great strides in the last
  two decades, to the point that their application in safety-critical embedded
  software is an undeniable success. Their extension to non-critical software is
  one of the notable forthcoming challenges. For example, C programmers
  regularly use inline assembly for low-level optimizations and system
  primitives. This usually results in rendering state-of-the-art formal
  analyzers developed for C ineffective.  We thus propose \approach/, the first
  automated, generic, \emph{verification-friendly} and trustworthy lifting
  technique turning inline assembly into semantically equivalent C code amenable
  to verification, in
  order to take advantage of existing C analyzers. Extensive experiments on
  real-world code (including \gmp/ and \ffmpeg/) show the
  feasibility and benefits of \approach/.
\end{abstract}

\begin{IEEEkeywords}
Inline assembly, software verification, lifting, formal methods.
\end{IEEEkeywords}
%

%------------------------------------------------------------------------

%\togglefalse{islongpaper}
\toggletrue{islongpaper}

\section{Introduction}
\label{sec:intro}

\paragraph{Context}
Formal methods for the development of high-safety software have made tremendous
progress over the last two
decades~\cite{DBLP:journals/fac/KirchnerKPSY15,DBLP:conf/sas/DelmasS07,DBLP:journals/cacm/GodefroidLM12,DBLP:conf/osdi/CadarDE08,DBLP:conf/lics/OHearn15,DBLP:conf/cav/BallBLKL10},
 with notable success in regulated safety-critical industrial areas such as
avionics, railway or energy.
Yet, the application of formal methods to more usual (non-regulated) software,
for safety or security, currently remains a scientific challenge. In particular,
extending the applicability from a world with strict coding guidelines and
disciplined mandatory validation processes to more liberal and diverse
development and coding practices is a difficult task.

\paragraph{Problem}
We consider here the issue of analyzing ``mixed code'', focusing on the use of
inline assembly in \corcpp/ code. This feature allows to embed assembly
instructions in \corcpp/ programs. It is supported by major \corcpp/ compilers
like \gcc/, \clang/ or \vs/, and  used quite regularly --- usually for
optimization  or to access system-level features  hidden
by the host language. For example, we estimate that {\bf 11}\% of Debian packages written in
C/C++ directly or indirectly depends on inline assembly, with chunks containing up
to 500 instructions, while {\bf 28}\% of the top rated C projects on GitHub
contains inline assembly according to Rigger et al.
\cite{Rigger:2018:AXI:3186411.3186418}. As a matter of fact,
{\it inline assembly is a common
  engineering practice in key areas such as cryptography, multimedia or drivers.}
However, {\it it is not supported by current
  state-of-the-art \corcpp/ program analyzers}, like KLEE \cite{DBLP:conf/osdi/CadarDE08} or
Frama-C \cite{DBLP:journals/fac/KirchnerKPSY15}, possibly leading to
incorrect or incomplete results. {\it This is a clear applicability issue for
  advanced code analysis techniques.}

Given that developing  dedicated analyzers from scratch is too costly, the usual
way of dealing with assembly chunks is to write either equivalent host code (e.g, \corcpp/)
or equivalent logical specification when available.
 But {\it this
  task is handled manually} in both cases, precluding regular analysis of large
code bases: manual translation is indeed time-consuming and
error-prone. The bigger the assembly chunks are, the bigger these problems~loom.

\paragraph{Goal and challenges}
{\it We address the challenge of designing and developing an automated and
  generic lifting technique turning inline assembly into semantically equivalent
  C code amenable to verification.}  The method should~be:
\begin{description}

\item[Verification-friendly] The produced code should allow {\em good enough}
  analyses in practice (informally dubbed \emph{verifiability}), independently
  of the underlying analysis techniques (e.g., symbolic execution
  \cite{King1976,DBLP:journals/cacm/CadarS13}, deductive verification
  \cite{Floyd67,DBLP:journals/cacm/Hoare69} or abstract interpretation
  \cite{DBLP:conf/popl/CousotC77});

\item[Widely applicable]  It should not be tied to a particular architecture,
  assembly dialect or compiler chain, and yet handle a significant subset
  of assembly chunks found in the wild;

 \item[Trustworthy]  The translation process should be insertable in a formal
   verification context without endangering
soundness: as such it should maintain exactly all
  behaviors of the mixed code, and provide a way to show this
  property.
\end{description}
\IEEEpubidadjcol

Verifiability alone is already challenging: indeed, straightforward lifting from
assembly to C (keeping the untyped byte-level view) does not ensure it as
standard C analyzers are not well equipped to deal with such low-level C code.

Scarce previous attempts do not fulfill all the
objectives above. Vx86~\cite{Maus2008} is tied to both the \x86/ architecture and
deductive verification, while the recent work by Corteggiani et
al.~\cite{DBLP:conf/uss/CorteggianiCF18} focuses on symbolic execution. None of
them addresses verifiability or trust.
At first sight, decompilation
techniques~\cite{DBLP-conf/icsm/CifuentesSF98,DBLP-journals/jpl/Cifuentes96,
  DBLP-journals/spe/CifuentesG95} may seem to fit the bill. Yet, as they mostly
aim at helping reverse engineers, correctness is not their main
concern. Actually, \emph{``existing decompilers frequently produce decompilation
  that fails to achieve full functional equivalence with the original
  program''}~\cite{SchuRuNoCiLo18}. Some recent works partially target this
issue:
Schwartz et al.~\cite{DBLP:conf/uss/BrumleyLSW13} do not {\em demonstrate} correctness (they
instead measure a certain degree of it via testing), while Schulte et
al.~\cite{SchuRuNoCiLo18} use a correct-by-design but intractable (possibly non-terminating)
search-based method. Again, none of them study verifiability.

\paragraph{Proposal}
We propose \approach/ (Taming Inline Assembly), the first automated, generic,  verification-friendly
and trustworthy lifting technique for inline assembly.
The main insight behind \approach/ is that by {\it focusing on inline assembly
  rather than arbitrary decompilation, we tackle a problem both more
  restricted (simple control-flow, smaller size)
  and better defined (interfaces with C code, no dynamic jumps), paving the way to
  powerful targeted methods}.~\approach/~relies~on~the~following~key~principles:

\begin{itemize}%[leftmargin=1em]

\item Recent binary-code lifters
  \cite{Brumley2011,DBLP:conf/cav/BardinHLLTV11,DBLP-conf/kbse/KimFJJOLC17}
  translating binary opcodes to {\it generic  low-level intermediate representations}
  (IR)  provide minimalist architecture-agnostic and well-tested IRs adapted to
  our goal;

\item While direct byte-level lifting severely hinders current C analyzers,
  {\it verifiability is enhanced by dedicated transformations} refining the raw original IR
  with C-like abstractions such as explicit variables, arithmetic data
  manipulation, structured control-flow, etc.;

\item Trust relies on {\it translation validation} \cite{DBLP:conf/pldi/Necula00} (validating each
  translation), a more tractable option than full translator validation, which reduces the trust base to a
  (usually simpler) {\it checker}. Here, this checker requires to prove program equivalence -- a
  notoriously hard problem\footnote{Recall that general software verification
    problems, including program equivalence, are undecidable. Yet, software
    verification tools do exist and have been proven useful in practice.}. We
  propose a {\it dedicated equivalence checking algorithm} tailored to our
  processing chain.

\end{itemize}

\paragraph{Contributions}
In summary, this paper makes the following contributions:

\begin{itemize}[]%[leftmargin=1em]

\item A new  cooperating toolchain  allowing formal verification of programs
  mixing inline assembly and C, based on an original
  combination of novel and existing components (\cref{sec:method-overview}),  addressing
  verifiability and trust issues;

\item  A new principled method lifting inline assembly to
  high-level C amenable to further formal analysis built upon 4 simplification
  steps (\cref{sec:lifting})
  countering clearly  identified threats to verifiability (\Cref{sec:motivation});

\item The automated validation of said method  to make
  the lifter trustworthy, via a new dedicated program equivalence
  checking algorithm taking advantage of our transformation process to achieve both
  efficiency and high success rate, with a limited trust
  base (\cref{sec:validation}) ;

\item Thorough experiments (\cref{sec:xps}) of a prototype implementation on
  real-world examples to show its wide applicability (all Debian
  GNU/Linux 8.11 \x86/ assembly chunks, some \arm/, \gcc/ and
  \clang/) and its substantial impact
  on 3 different verification techniques on samples from \gmp/, \ffmpeg/, \alsa/
  and \libyuv/.

\end{itemize}

\paragraph{Discussion}

This work targets assembly chunks as found in real-world programs: we lift and
validate 76\% of all assembly chunks from Linux Debian
8.11~(\cref{tab:wa-debian}) and benefit a range of state-of-the-art verification
tools and techniques~(\cref{sec:verification}).
Still, system and floating-point instructions are currently considered
out-of-scope.
Especially, floats are not tackled here since handling them well is a
challenge in itself for the whole toolchain (lifter, solver, verifier) --- see
the extended discussion in~\cref{sec:discussion}.
Also, \approach/'s implementation targets C since this is the
principal language used for low-level programs, but the method itself would work
unchanged on similar imperative languages, like LLVM.
Finally, though some prior work has addressed code lifting for verification,
it is worth noting that {\it verifiability} has never been explicitly addressed so far.

\section{Context and motivation}
\label{sec:motivation}

\tikzstyle{bb}=
          [draw, rectangle, ultra thick, minimum width=100, minimum height=25]
\tikzstyle{edge}=[very thick, ->]

\newcommand{\vid}[1]{{\sf #1}}
\newcommand{\vin}[1]{\vid{\textcolor{darkblue}{#1}}}
\newcommand{\vout}[1]{\vid{\textcolor{darkred}{#1}}}

\begin{figure}[htbp]
  %\vspace*{-1cm}
  \hspace*{.5cm}
  \centering
  \begin{subfigure}[t]{\linewidth}
    \hspace*{-20pt}\includegraphics{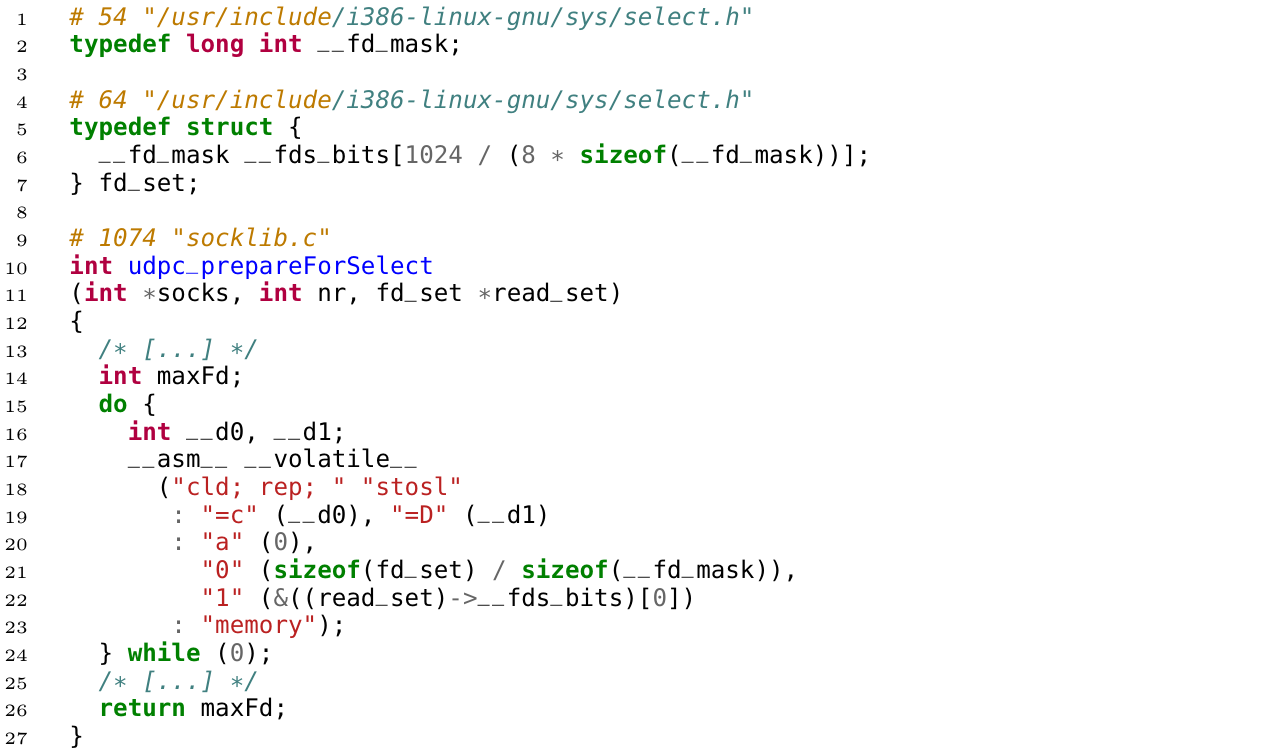}

    %\vspace*{-.4cm}
    \caption{Original version}
    \label{fig:motivation_example_asm_part}
  \end{subfigure}
  % \vspace*{-1cm}
  \vspace*{.4cm}

\begin{subfigure}[b]{\linewidth}
  {
    %\vspace*{-.2cm}
    \setlength{\tabcolsep}{1pt}
    \centering
    % \resizebox{3 \linewidth / 4}{!}{
    \begin{adjustbox}{width=3 \linewidth / 4,center}
      \begin{tikzpicture}[
          edge/.style={->, black, shorten >= 2pt, shorten <= 2pt},
          bb/.style={rectangle, draw, solid, rounded corners=1mm, minimum size=.7cm,
           inner sep=5pt},
        ]
        \coordinate (entry);
        \node[right = .1 of entry] {in};

        \node[bb, anchor = north] (a) at ($(entry) + (right:1.4) + (down:0.75)$)
             {\begin{tabular}{rcl}
                 \vout{eax} & $\leftarrow$ & 0x00000000 \\
                 \vout{ecx} & $\leftarrow$ & \cinl{sizeof(fd_set) / sizeof(__fd_mask)} \\
                 \vout{edi} & $\leftarrow$ & \lstinline[language=c, basicstyle=\footnotesize]$&((read_set)->__fds_bits)[0]$ \\
                 \vout{df}  & $\leftarrow$ & 0 \\
             \end{tabular}};

        \node[bb, anchor = north] (b) at ($(a.south) + (left:1.4) + (down:0.75)$)
             {\begin{tabular}{rcl}
                 \multicolumn{3}{c}{if \vin{ecx} = 0 then \textcolor{darkblue}{\sf break}} \\
             \end{tabular}};

        \node[bb, anchor = west, right = 0.5 of b] (c)
             {\begin{tabular}{rclclcl}
                 \vout{@[}&\vin{edi}&\vout{]$_4$} & $\leftarrow$ & \multicolumn{3}{l}{\vin{eax}} \\
                 &\vout{edi}& & $\leftarrow$ & \vin{df} & ? & \vin{edi} $-$ 4 \\
                 &&           &              &          & : & \vin{edi} $+$ 4 \\
                 &\vout{ecx}& & $\leftarrow$ & \multicolumn{3}{l}{\vin{ecx} $-$ 1} \\
             \end{tabular}};

        \coordinate[below = .75 of b]  (exit);
        \node[right = .1 of exit] {out};

        \draw[edge, thick] (entry) -- ($(a.north) + (left:1.4)$);
        \draw[edge] ($(a.south) + (left:1.4)$) -- (b);
        \draw[edge] ([yshift=-5]b.east) -- ([yshift=-5]c.west);
        \draw[edge] ([yshift=5]c.west) -- ([yshift=5]b.east);
        \draw[edge, thick] (b) -- (exit);

      \end{tikzpicture}
    \end{adjustbox}
    % }
  }
  \vspace*{-.5cm}
  \caption{Low-level semantics}
  \label{fig:motivation_example_dba_part}
  \end{subfigure}
  %\vspace*{-1.2cm}

  \hspace*{.5cm}
  %\vspace*{-.6cm}
  \begin{subfigure}[t]{\linewidth}

    \vspace*{.2cm}
    \hspace*{-20pt}\includegraphics{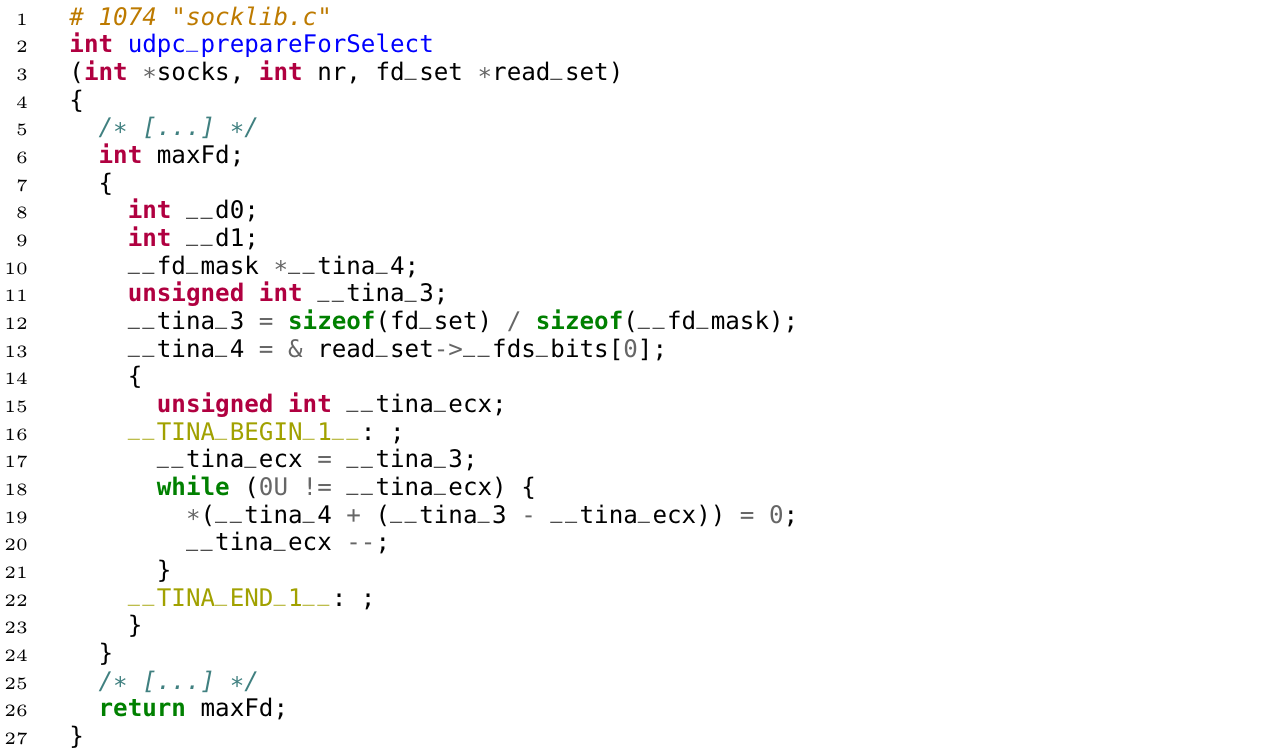}

    %\vspace*{-.4cm}
    \caption{\approach/-generated version}

    \label{fig:motivation_example_tina_result}
  \end{subfigure}
  \bigskip
  \caption{Running example}
  \label{fig:motivation_example}
  \label{fig:running_example}
\end{figure}

Consider the code snippet of \cref{fig:motivation_example_asm_part}  extracted
from \udpcast/ sources. It consists of the \x86/ assembly code itself (here: \cinl{"cld; rep stosl"})
together with a {\it specification} linking C variables to registers and declaring inputs,
outputs and clobbers (i.e., registers or memory cells possibly modified by the assembly chunk).
  The compiler, upon encounter of such an (extended) assembly
chunk, may use this { specification}
-- for example during register allocation. However, it is fully
blind to the rest of the information (e.g., mnemonics) and will forward the chunk
{\em as is} until code emission.

\paragraph{Annotation} The code in \cref{fig:motivation_example_asm_part} is
suffixed by a specification, written in a concise constraint language
(\gcc//\clang/ syntax), in zones
separated by '\cinl{:}' (lines 16-23):

\begin{itemize}%[leftmargin=1em]
\item It first describes allocation constraints for output variables:
  \begin{enumerate}[leftmargin=2em,start=0,label={\bf "\arabic*"}.,ref="\arabic*"]%[label=o\arabic*.,ref=o\arabic*]
  \item \label{cst:d0}
    \cinl{"=c" (__d0)} specifies that variable \cinl{__d0}
    should be assigned to register \cinl{ecx};
  \item \label{cst:d1}
    \cinl{"=D" (__d1)} specifies that variable \cinl{__d1}
    should be assigned to register \cinl{edi};
  \end{enumerate}
\item Then, lines 20-22 detail inputs: \cinl{"a" (eax)} holds \cinl{0},
\cinl{sizeof(fd_set) / sizeof(__fd_mask)}
is held in the register described in \ref{cst:d0} (\cinl{ecx})
and the one described in \ref{cst:d1}
  (\cinl{edi}) holds \cinl{&((read_set)->__fds_bits)[0]};

\item Finally, the whole memory (\cinl{"memory"})
  can be assigned. This basically tells the compiler to flush its memory cache
 before entering the chunk.
\end{itemize}

\paragraph{Informal semantics} The code \cinl{"cld; rep stosl"}
has the following informal semantics (\cref{fig:motivation_example_dba_part}):
put the direction flag \cinl{df} to \cinl{0}, then fill \cinl{ecx} double
words from the \cinl{edi} pointer with the value from \cinl{eax}. Intel's manual
\cite{intelv2} explains that  \cinl{df} drives the sign of the increment: when \cinl{df} is
\cinl{0}, the sign is positive. \approach/ produces the code in
\cref{fig:motivation_example_tina_result}: the loop from the informal semantics
is there, but the lifter optimized away (see~\cref{sec:code-optim})
elements like \cinl{df}, \cinl{eax} or \cinl{edi}.

\paragraph{Running the analyzers}
If we try to run industrial-strength C code analyzers on this code, we observe erratic behaviors:
KLEE \cite{DBLP:conf/osdi/CadarDE08}  stops with an error message; Frama-C, on the other hand, warns that
{\sf \footnotesize 'Clobber list contains "memory" argument. Assuming no side-effect
  beyond those mentioned in output operands'}. This message is clear but the
behavior incorrect:
the keyword \cinl{"memory"} stipulates that all memory may be assigned but
Frama-C simply ignores it. This small example shows that a single line of
assembly may throw off these tools.
Of course, one may manually rewrite  the chunks into semantically
equivalent C code, then use C analyzers, but this is error-prone and not scalable.
\textbf{With \approach/}, we are able to automatically generate the code of
\cref{fig:motivation_example_tina_result}, illustrative of our code transformations (see
\cref{sec:lifting}), and automatically validate it \textit{(trustworthy)}.
We can then formally show  with
Frama-C~\cite{DBLP:journals/fac/KirchnerKPSY15}, using abstract interpretation or
deductive verification, that the code indeed verifies the informal semantics
laid out before \textit{(verification-friendly)}.

\paragraph{Identified threats to verifiability}
Straightforward lifting from assembly to C (keeping the untyped byte-level view)
does not ensure verifiability, as standard C analyzers are not well equipped to
deal with such low-level C code. For example we cannot prove with Frama-C that a basic lifting of
\cref{fig:motivation_example_asm_part} meets its
specification\begin{long-paper}\phantom{ }(cf.~Appendix~\ref{sec:example-ext})\end{long-paper}.
We identify 3 main threats to
verifiability:
\begin{enumerate}[series=threats,leftmargin=2em,label={\bf T\arabic*.},ref={\bf T\arabic*}]

\item \label{threat:data}
  Low-level data: explicit flags -- including overflows or carry,
  bitwise operations (masks), low-level comparisons, byte-level memory;

\item \label{threat:vars}
  Implicit variables: variables in the untyped byte-level stack, packing
  of separate logical variables inside large-enough registers;

\item \label{threat:loop}
  Implicit loop counters/index: structures indexed by loop counters  at high-level are split into multiple low-level
  computations where the link between the different logical elements is
  lost.
\end{enumerate}

Experiments in \cref{sec:verification} demonstrate that a straightforward
encoding (\textsc{Basic}) fails to get the best of any analysis -- symbolic
execution, abstract interpretation, or deductive verification.

\paragraph{Properties of inline assembly}
\approach/ exploits the following properties, specific
to inline assembly:
\begin{enumerate}[leftmargin=2em,label={\bf P\arabic*.},ref={\bf P\arabic*}]
\item \label{prop:simple}
  The control flow structure is limited: only a handful of conditionals and
  loops, hosting up to hundreds of instructions;
\item \label{prop:annotated}
  The interface of the chunk with the C code is usually given:
  programmers annotate chunks with the description of its inputs,
  outputs and clobbers with respect to its C context;
\item \label{prop:context} Furthermore, the chunk appears in a C context, where
  the types, and possibly more, are known: this kind of information is sought
  after in decompilers, using heuristics, whereas we only need to propagate it here.
\end{enumerate}

All in all, the above points show that lifting assembly chunks   is actually an
interesting sub-problem of general decompilation, both simpler and richer in
information and thus significantly more amenable to
overall success.

\section{Background}
\label{sec:background}

\subsection{Inline Assembly}
\label{sec:inline-assembly}

We  focus here on inline assembly in \corcpp/ code as supported by \gcc/ and
\clang/.  MASM (Microsoft Macro Assembler) has a different syntax but works
similarly.

\begin{figure}[!htbp]
  \centering
  %\vspace*{-1cm}
  \subcaptionbox{Basic version\label{fig:basic_asm_example}}
                {\begin{minipage}[b]{0.20\textwidth}
                    \includegraphics{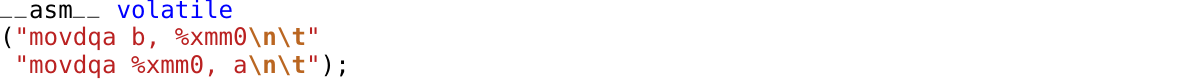}
                \end{minipage}}
  \subcaptionbox{Extended version\label{fig:extended_asm_example}}
                {\begin{minipage}[b]{0.20\textwidth}
                    \includegraphics{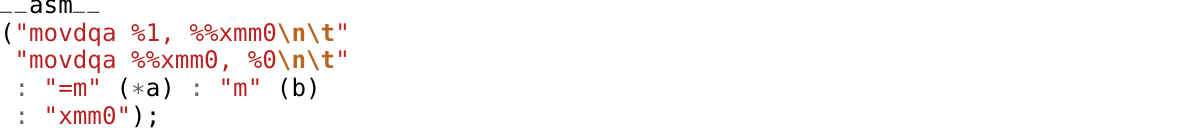}
                \end{minipage}}
  \caption{Assembly chunks: basic \& extended versions}
  \label{fig:asm_chunk}
\end{figure}

Assembly chunks in the GAS syntax of \gcc/ have two flavors:
basic and (recommended) extended (see \cref{fig:asm_chunk}).  {\it Basic
assembly} (\cref{fig:basic_asm_example}) allows the insertion of assembly
instructions anywhere in the code. They will be emitted \emph{as is} during the
production of the assembly file. In this case, compilers \emph{assume} the chunk has no
effect on its C scope, {\it preventing safe interactions between assembly and C
code} -- yet  that does not stop developers from using it when the implicit
context {\it looks} safe. In \cref{fig:basic_asm_example}, it is
{\it implicitly}  assumed
that no optimization will
occur on global variables \cinl{a} and \cinl{b} and that \cinl{xmm} registers are not
used by default.

\paragraph{Extended assembly}
{\it Extended assembly} allows in addition the description of the interactions with C
through its inputs, outputs and clobbers (i.e., registers or memory cells whose value is rewritten by the chunk).

Such annotations work like a \cinl{printf} string format, as shown in
\cref{fig:extended_asm_example}: some assembly operands may be replaced by
placeholders referring to a list of C operands.
The syntax requires binding C operands to
location constraints, as in \cref{fig:motivation_example_asm_part,fig:extended_asm_example}.
Constraints may also specify more than one location, and let the compiler choose
the best way to place this operand. Common placement constraints include
\cinl{r} to bind to a general register;
\cinl{m} to bind to a memory address;
\cinl{i}  to an immediate value; and \cinl{g}
which means "\cinl{r}, \cinl{m} or \cinl{i}".
Operands may be read-only (for inputs) or write-only (for outputs) with the {\sf =} modifier.
A read-write operand is created either by linking an input
to the same location as an output (\cinl{"0" (sizeof(fd_set) / sizeof(__fd_mask))}
in \cref{fig:motivation_example_asm_part}) or by using the
{\sf +} modifier instead of
{\sf =}. Without special modifiers, compilers assume read-only
operands are consumed before write-only operands are produced, so that these may
share the same locations.
The clobber list may also contain keywords like \cinl{"memory"} (arbitrary memory
cells may be read and/or written) and \cinl{"cc"} (conditional flags will be
changed).

The specification of inputs, outputs and clobbers stands as a contract between
the chunk and the compiler. Compilers are totally blind to what actually happens
inside the chunk, relying on the contract, and will not warn about mistakes inside
the chunk. Forgetting to list an input or a clobber is an easy mistake that can result
in code which does not behave as expected.

\paragraph{Adoption}
The use of inline assembly is
 pretty widespread --  we estimate that 11\% of Debian packages written in
C/C++ directly or indirectly depends on inline assembly. It includes major projects
like \gmp/ and \ffmpeg/ --- Rigger et al. \cite{Rigger:2018:AXI:3186411.3186418}
actually reports that 28\% of the top rated C projects on GitHub uses inline assembly.
We further estimate that 75\% of the chunks found in Debian Jessie 8.11 (used in
\cref{sec:applicability}) serve an optimizing purpose, with an average size of
approximately 10 instructions, and up to 341.  Inline assembly is often used in conjunction
with C macros or inlineable functions to be specialized by the compiler at each
location.

\subsection{Binary-code lifters}
\label{sec:binary-lifter}

Binary-code lifters are the cornerstones of modern binary-level
analyzers. They are used to abstract the different binary Instruction Set
Architectures (ISA) and formats into
a single intermediate representation (IR)
\cite{Brumley2011,DBLP:conf/cav/BardinHLLTV11,DBLP-conf/kbse/KimFJJOLC17}.
We rely on the IR of \binsec/ \cite{Djoudi2015}, called DBA --- other
IRs are similar. Its syntax is shown in \cref{fig:dba_ir}.

\newcommand{\te}[1]{{\sf{#1}}}
\begin{figure}[htbp]
  \centering\scriptsize
  \begin{tabular}{rcl}
    \te{inst} & {\sf :=} & \te{lv} $\bm\leftarrow$ \te{e} | {\sf goto e}
                          | {\sf if} \te{e} {\sf then goto e}  {\sf else goto} \te{e} \\
    \te{lv} & {\sf :=} & {\sf var} | {\sf @[}\te{e}{\sf ]}$_{n}$ \\
    \te{e} & {\sf :=} & {\sf cst} | \te{lv}
                      | \te{unop} \te{e} | \te{binop} \te{e} \te{e}
                      | \te{e} {\sf ?} \te{e} {\sf :} \te{e} \\
    \te{unop} & {\sf :=} & ${\bm \neg}$ | ${\bm -}$ | {\sf uext$_{n}$}
                          | {\sf sext$_{n}$}    | {\sf extract$_{i .. j}$} \\
    \te{binop} & {\sf :=} & \te{arith} | \te{bitwise} | \te{cmp}
                           | {\sf concat} \\
    \te{arith} & {\sf :=} & ${\bm +}$ | ${\bm -}$ | ${\bm \times}$
                           | {\sf udiv}    | {\sf urem}
                           | {\sf sdiv}   | {\sf srem} \\
    \te{bitwise} & {\sf :=} & ${\bm \land}$ | ${\bm \lor}$ | ${\bm \oplus}$
                             | {\sf shl} | {\sf shr} | {\sf sar} \\
    \te{cmp} & {\sf :=} & ${\bm =}$ | ${\bm \neq}$
                         | $>_{u}$   | $<_{u}$
                         | $>_{s}$         | $<_{s}$
  \end{tabular}
  \caption{Low-level IR for binary code}
  \label{fig:dba_ir}
\end{figure}

DBA is a minimalist language, comprising only two types of
elements (bitvector values and memory) and three instructions: assignments,
jumps and conditionals. Yet, this is  enough to encode the
functional semantics of major ISAs -- including \x86/ and \arm/.

Binary lifters provide specialized decoders for supported
architectures, in the same spirit that a compiler has one code emitter per
supported architecture. Lifters are then used in disassembly algorithms to
(try to) recover the semantics of the binary program. We use them to disclose the
semantics of compiled assembly chunks.

\section{Taming Inline Assembly: an overview}
\label{sec:method-overview}

\approach/ lifts inline assembly to semantically equivalent C taking advantage
of properties \ref{prop:simple}--\ref{prop:context}.  This original  process consists mainly of
two (new) phases: {\em verification-friendly lifting} and {\em validation}, detailed respectively in
\cref{sec:lifting,sec:validation}. First, let us discuss the overall
approach, as schematized in \cref{fig:flow_overview}.

\newcommand{\trusted}[1]{\textcolor{highlight}{\textbf{#1}}}
\begin{figure*}[htbp]
  \centering
  \resizebox{.8\textwidth}{!}{
  \begin{tikzpicture}[
      arr/.style={->, black, shorten >= 2pt, shorten <= 2pt},
      box/.style={rectangle, draw, solid, rounded corners=1mm,minimum size=.7cm,
      inner sep=2pt},
      dec/.style={diamond,draw,solid,inner sep=3pt},
      tr/.style={font=\scshape\footnotesize},
      title/.style={font=\large\scshape,fill=bggrey}
    ]

    \node (init) {};
    \node (c) at ($(init) + (right:1.5)$) {\sf C};
    \node (plus) at ($(c) + (right:.4)$) {+};
    \node (asm) at ($(plus) + (right:.6)$) {\asm/};
    \node[fit=(c)(asm), draw, inner xsep=10pt, inner ysep=4pt, rounded corners=1mm] (c+asm) {};

    \node[box] (exe1) at ($(c+asm) + (right:4.5)$) {Object code};

    \node[box] (dba1) at ($(exe1) + (right:4.5)$) {IR};
    \node[tr] (opt) at ($(dba1) + (up:1)$) {Transformations};

    \node[box, ultra thick, fill=bggrey] at ($(dba1) + (right:4.5)$) (cport) {\sf C};
    \node (out) at ($(cport) + (right:1)$) {};

    \node[box] at ($(cport) + (down:1.5) + (left:3.3)$) (exe2)  {Object code};
    \node[box] (dba2) at ($(exe2) + (down:2)$) {IR};
    \draw[arr] (asm) -- node[tr,above] {Compilation} node[tr,below] {+ \trusted{Debug}} (exe1);
    \draw[arr] (exe1) -- node[tr,above] (irport) {\trusted{IR lifting}} (dba1);

    \node[box] at (irport |- dba2) (dba_orig)  {IR};
    \draw[arr] (dba1) -- node[tr,above] (insert) {Insertion} (cport);
    \draw[arr] (irport) -- (dba_orig);
    \draw[arr] (cport) --  (cport |- exe2) --
    node[tr,above] {\trusted{Compilation}}
    node[tr,below] (d2) { +  \trusted{Debug}}
    (exe2);
    \draw[arr] (exe2) -- node[tr,right] (pir) {\trusted{IR Lifting}} (dba2);
    \draw[<->, black, shorten >= 2pt, shorten <= 2pt] (dba_orig) --
          node[tr,above] (eq) {\trusted{Equivalence}} node[below] {\good / \bad?} (dba2);
    \draw[arr, thick] (init) -- (c);
    \draw[arr, thick] (cport) -- (out);

    \node at ($(c) + (up:1.8)$) (cup) {};
    \draw[arr] (c) -- (cup.center) -- node (c2c) {} (cup -| cport) -- (cport);

   \path (dba1) edge[arr, loop] (dba1);
    \node[fit=(c+asm)(dba1)(c)(insert)(opt)(c2c), draw, dotted, inner ysep=8pt, inner xsep=8pt, rounded corners=1mm]
    (translate) {};
    \node at ($(c+asm)+(up:1.5)$) (t) {};
    \node[title ] at ($(translate)+(up:.5)$) {Lifting};

    \node[fit=(dba_orig)(dba2)(exe2)(pir)(d2),draw, dotted, inner ysep=5pt,
    inner xsep=10pt, rounded corners=1mm] (validate) { };
    \node[title] at ($(eq) + (up:.8)$) (val) {Validation};
    \node at ($(val) + (left:7)$) {
      \begin{minipage}{0.3\linewidth}
        \trusted{Highlighted elements} discussed in
        \cref{sec:trustbase} about the trust base.
      \end{minipage}
};
  \end{tikzpicture}}
  \caption{Overview of \approach/}
  \label{fig:flow_overview}
  \vspace*{-5pt}
\end{figure*}

\paragraph{Compilation}
We compile the source code for the target architecture {\em with debug
  information}. Since we control code compilation, we also include all
contextual data that can help to reconstruct C code, e.g., variable names
and types.

\paragraph{Initial low-level IR lifting [genericity]}
We now start the translation per se, by lifting the code back to the IR level.
The use of binary code may seem gratuitous at first
sight. This is however the best place to start working, since assembly chunks
are totally instantiated and embedded in their context --- register names and
memory locations have been resolved by the compiler. Debug information here allows to
locate the assembly chunk in the compiled code.

\paragraph{Transformation into high-level C [verifiability]}
We then lift the IR back to C, through a combination of dedicated  passes
aiming at refining the low-level IR with high-level information  (\cref{sec:lifting}).
The end result is a C-only code where assembly chunks
have been substituted by a lifted C code amenable to verification.
{\it This step is original.}

\paragraph{Validation [trust]}
The validation phase starts by recompiling the pure C code, {\em without
  optimization} in order to preserve the code structure --- our validation
technique depends on it. We locate the binary code corresponding to the lifted
code once more, and get back its IR representation. We now possess two distinct
IR pieces: this one and the one from the first compilation. We will aim to prove their semantic
equivalence in~\cref{sec:validation}. {\it This step is original.}

\smallskip We have \textit{\textbf{implemented}} {\it a prototype of \approach/} leveraging
existing tools: Frama-C~\cite{DBLP:journals/fac/KirchnerKPSY15} for C source
code manipulation (parsing, localization, C injection),
\binsec/~\cite{Djoudi2015,DBLP:conf/wcre/DavidBTMFPM16} (IR
lifting~\cite{DBLP:conf/cav/BardinHLLTV11}, SMT solvers
integration~\cite{DBLP-reference/mc/BarrettT18}), and the DWARF~\cite{dwarf5}
debug format to pass information to binaries with the compiler.

\section{From low-level IR to high-level C}
\label{sec:lifting}
\label{sec:code-optim}

The goal of this lifting phase is to recover {\em verifiable} C code
preserving the semantics of the original assembly chunk.
The transformations at IR level mitigate the identified threats to
verifiability (\cref{sec:motivation}), and reinforce each
other (\cref{sec:xps}).

\paragraph{Type verification \& propagation}
\label{sec:code-lifting}
To lift assembly code back to C, chunk operations on bitvectors and memory need
to be mapped to C operations on integers (signed/unsigned) and pointers.
To this end, we propagate types from the interface into the IR operations.
IR types can either be addresses (typed pointers) or values (signed or unsigned, with an
associated size). Type information is further synthesized using forward
propagation and constraints imposed on operands by low-level operations.
This step also guarantees that inputs' and outputs' types are respected.
The lifter gives concrete C types using the
type size information from DWARF.

\paragraph{High-level predicate recovery  (threat~\ref{threat:data})}
\label{sec:hlp-recovery}
Low-level conditionals use flags --- {\sf zero} (\cinl{zf}), {\sf
  sign} (\cinl{sf}), {\sf carry} (\cinl{cf}) or {\sf
  overflow} (\cinl{of}) --- set by previous instructions. In most situations, they have
little meaning on their own and the way they are computed hampers understanding
the purpose of the condition. This pass applies Djoudi
et al.'s recent technique~\cite{Djoudi2016} based on semantic equivalence proved by SMT solvers. It
substitutes the low-level condition, built on flags, by a more readable arithmetic
comparison. For example, this phase recovers \cinl{if (ecx + 1 > 1) goto label;}
instead of \cinl{if (zf == 0 && sf == of) goto label;} from the assembly snippet
\cinl{"decl ecx; jg label;"}.

\paragraph{Register unpacking (threat \ref{threat:vars})}
\label{sec:unpacking}
Assembly chunks often contain optimizations exploiting data level parallelism
in order to use the full capacity of the hardware by packing multiple value
inside a bigger one fitting inside a machine register. For instance, loading 4
(byte) characters inside an integer is more efficient than doing four smaller
loads.  The concept has been exacerbated with {\em Single Instruction Multiple
  Data} extensions, providing vectorized registers up to 512-bits.
The issue here is that such {\it packed code}  has very low-level semantics (masks, shifts, etc.).
Our novel \emph{register unpacking} method
uncovers the independent variables stored in a container, thus preventing packed
arithmetic from destroying the abstractions of the analyzers.
The method amounts to splitting registers into independent variables, whose size
depends on the uncovered usage, rewrite the code
accordingly and then clean up unused
variables and code, and rebuild higher-level chunks through dedicated
simplifications.
The principle is the following:  if a subpart of a variable is read in the code
(e.g., {\small\sf extract$_{0 ..  15}$} \cinl{eax}), then

  this subpart is likely to correspond to a logical entity. So we generate a
  fresh variable for this entity, receiving the restricted value, and replace
  each such extraction by this new variable.
  To avoid the need for a fixpoint until every variable extraction
  is replaced,
  we perform the replacement eagerly, in 3 steps:
\begin{enumerate}
\item A forward pass where each assignment of $8\times2^k$ bits is split into
multiple fresh assignments of $8\times2^i$ bits where $i \leq k$ (for instance,
\cinl{eax} will be split into \{\{\cinl{al}, \cinl{ah}, \cinl{eax_16_23},
\cinl{eax_24_31}\}, \{\cinl{ax}, \cinl{eax_16_31}\}, \{\cinl{eax}\}\};

\item At the same time, each variable restriction {\sf extract$_{i .. j}$}
  \cinl{var} corresponding to one of the newly generated variables is replaced
  by this new variable;

\item A final pass of dead code elimination removes each unused
  freshly generated variable.
\end{enumerate}

Note that subparts may overlap with each other (for instance, \cinl{al},
\cinl{ax} and \cinl{eax} share common parts) but we found that most of the time,
only one of them survives the final step. Thus, the size of the produced code does not increase much in the end.

\medskip

  Finally, we also rely on the fact that expression propagations together with
  concatenation-extraction simplification will automatically reconstruct
  bigger sized variables from concatenation of smaller sized ones
  (e.g., \cinl{ax} half-word from \cinl{al} and \cinl{ah} bytes).

\begin{figure}[!htbp]
  \begin{subfigure}[t]{.45\linewidth}
    \vspace*{-2.7cm}
    \includegraphics{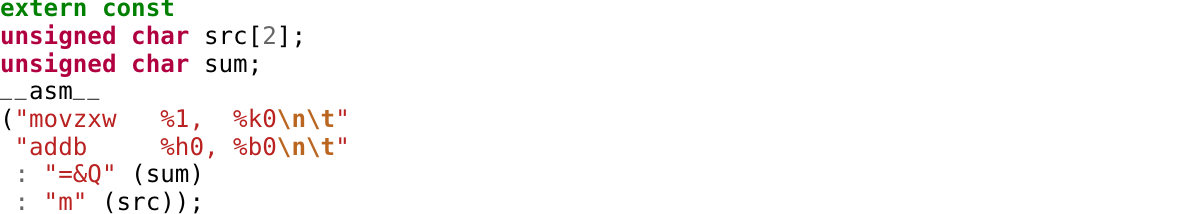}
        \vspace{-.5cm}
    \caption{Source}
    \label{fig:src_unpacking}
  \end{subfigure}
  \begin{subfigure}[t]{.5\linewidth}
    \includegraphics{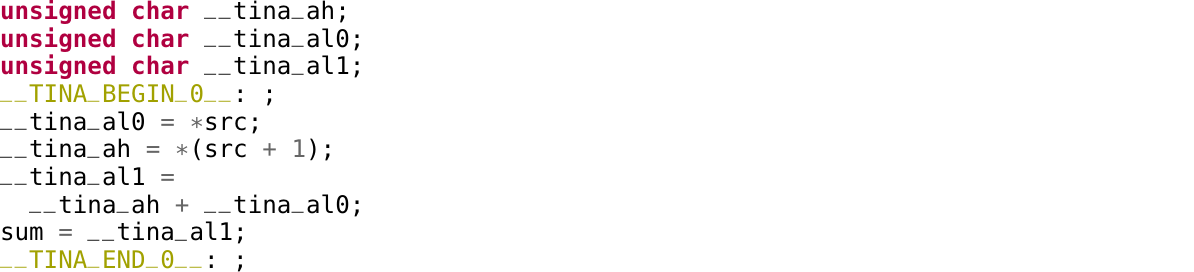}
        \vspace{-.5cm}
    \caption{Lifting {\em with} unpacking}
    \label{fig:optim_unpacking}
  \end{subfigure}
  \begin{subfigure}[t]{\linewidth}
    %\vspace{-1.2cm}
    \includegraphics{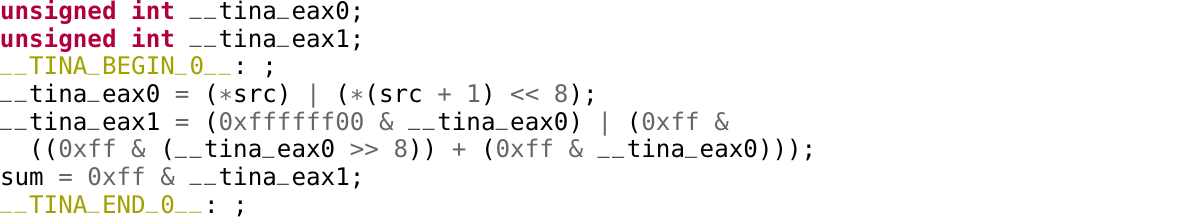}
    \vspace{-.5cm}
    \caption{Lifting {\em without} unpacking}
    \label{fig:optim_no-unpacking}
  \end{subfigure}
  \caption{Register unpacking}
  \label{fig:optim_unpacking_cmp}
\end{figure}

In
\Cref{fig:optim_unpacking_cmp},
the chunk loads two {\sf char} in a
register before adding them, using the \cinl{h} and \cinl{l} prefixes to access
them. Without register unpacking, the lifter uses
bitmasking~(\cref{fig:optim_no-unpacking}), making the code more complex than
its clear initial intent~(\cref{fig:optim_unpacking}).

\paragraph{Expression propagation (threats~\ref{threat:data} and
  \ref{threat:vars})} \label{sec:expr-propagation}
We draw
inspiration from compiler optimization techniques to devise a novel dedicated
simplification mechanism geared toward our needs.  In particular, we can afford
very aggressive simplifications (small code size w.r.t. standard compilation
setting) but we have to address particular kinds of low-level instructions
(coming from IR translation).  Our method originally combines \emph{eager
  expression propagation} coupled with \emph{dedicated (low-level)
  simplifications} and \emph{a posteriori control} to revert fruitless
propagations
  -- when no simplification rule has been triggered.

\emph{Eager expression propagation}  relies on the idea that more expression propagation raises
more opportunity for further simplifications by dedicated rules. Yet, systematic propagation can
yield an exponential blowup of the code under analysis rather than the desired simplifications.
To mitigate this problem we propose eager propagation coupled with a posteriori
control to revert fruitless propagation. The algorithm works as follows:

\begin{itemize}
\item As a preliminary step, a data flow analysis collects all symbolic values (terms) associated to each pair
(name, program point) used in the IR code;

\item First, we {\it unconditionally propagate} symbolic values in a first pass but save a {\it reverse map}
for each propagated expression (in case the propagation is not fruitful);

\item Second, we expect simplification rules (described below) to {\it simplify} the whole expression;

\item Third, we identify expressions not yet simplified (by syntactically
  comparing the terms before and after simplification) and revert back the
  propagation on such case thanks to the reverse map (\textit{a posteriori
    control});

\item Finally, we {\it cleanup} the code by filtering out unused variables, dead branches and dead code.
\end{itemize}

Regarding \emph{simplification rules},
we use a mixture of  standard and dedicated
simplification rules -- standard for typical integer-level properties and
dedicated for more low-level aspects. Here is a representative (incomplete)
subset of these rules
\begin{long-paper}
  -- see Appendix~\ref{sec-expression-propagation-ext} for a complete set --\phantom{ }
\end{long-paper}
where $|x|$ denotes the size of
the expression $x$, $\diamond$ any binary operator, $C$ a condition ($|C| = 1$), $k$ is a constant.

\setlength{\abovedisplayskip}{1pt}
\setlength{\belowdisplayskip}{1pt}
\setlength{\abovedisplayshortskip}{1pt}
\setlength{\belowdisplayshortskip}{1pt}

\begin{itemize}

\item associativity-commutativity re-ordering:
  \begin{equation*}
    x + 1 + a \hookrightarrow a  + x + 1
  \end{equation*}

\item constant propagation (modular arithmetic):
  \begin{gather*}
    10 + 5 \hookrightarrow 15,\quad 10 \times 2 \hookrightarrow 20
  \end{gather*}

\item standard  algebraic simplifications (identity, neutral, absorbing and inverse elements, etc.):
  \begin{gather*}
    x + 0 \hookrightarrow x,\quad x \times 1 \hookrightarrow  x,\quad
    x \times 0 \hookrightarrow  0,\quad x - x \hookrightarrow  0 \\
    x \lor 0 \hookrightarrow x,\quad x \land 1 \hookrightarrow  x,\quad x \land
    x \hookrightarrow x, \quad x \oplus x \hookrightarrow 0
  \end{gather*}

\item ternary expression simplification:
  \begin{align*}
    C \ ?\ x : x &\hookrightarrow x,& \neg C \ ?\ x : y &\hookrightarrow C \ ?\ y : x \\
    true \ ?\ x : y &\hookrightarrow x,& false \ ?\ x : y &\hookrightarrow y \\
    C \ ? \ true : false &\hookrightarrow C,& C \ ? \ false : true &\hookrightarrow \neg C
  \end{align*}

\item ternary expression development:
  \begin{align*}
    x \diamond (C \ ? \ y : z) &\hookrightarrow C \ ? \ x \diamond y : x \diamond z \\
    (C \ ? \ w : x) \diamond (C \ ? \ y : z) &\hookrightarrow
                                               C \ ? \ w \diamond y : x \diamond z
  \end{align*}

\item two-complement arithmetic abstraction:
  \begin{align*}
    \neg x + 1 &\hookrightarrow - x \\
    \text{extract}_{|x| - 1}(x) &\hookrightarrow x <_{s} 0
  \end{align*}
  \begin{align*}
    \text{uext}_{n}(C) - 1 &\hookrightarrow C \ ? \ -1_{n} : 0_{n}\\
    \text{sext}_{n}(C) &\hookrightarrow C \ ? \ -1_{n} : 0_{n}
  \end{align*}

\item concatenation:
  \begin{align*}
    \text{uext}_{|x|+|y|}(x) \vee \text{concat}(y, 0_{|x|}) &\hookrightarrow \text{concat}(y, x) \\
    \text{uext}_{|x| + k}(x) \mathbin{\text{shl}} k &\hookrightarrow \text{concat}(x, 0_{k}) \\
    \text{concat}(0_{k}, x) &\hookrightarrow \text{uext}_{k + |x|}(x)
  \end{align*}

\item extraction-concatenation simplification:
  \begin{align*}
    \text{extract}_{0 .. |x| - 1}(x) &\hookrightarrow x \\
        \text{concat}(\text{extract}_{i .. j}(x), \text{extract}_{j + 1 .. k}(x))
    &\hookrightarrow \text{extract}_{i .. k}(x)
  \end{align*}
    \begin{align*}
    \text{extract}_{i .. j}(\text{concat}(x, y))_{\ \mathbin{\text{when}}\ j < |y|}
                                     &\hookrightarrow \text{extract}_{i .. j}(y) \\
    \text{extract}_{i .. j}(\text{concat}(x, y))_{\ \mathbin{\text{when}}\ |y| \le i}
                                     &\hookrightarrow \text{extract}_{i - |y|
                                       .. j - |y|}(x)
    \end{align*}

\end{itemize}

\begin{figure}[htbp]
  \begin{subfigure}[t]{\linewidth}
    \centering
    \includegraphics{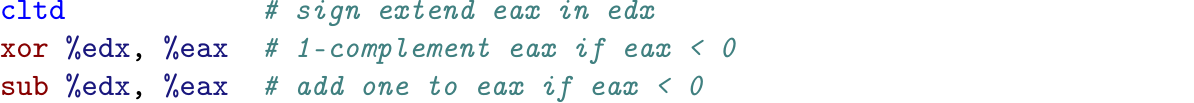}
    \caption{Branchless absolute value implementation}
    \label{fig:expr-propagation-src}
  \end{subfigure}

  \medskip

  \begin{subfigure}[t]{\linewidth}
    \centering
    \resizebox{\linewidth}{!}{
    \setlength{\tabcolsep}{.4em}
    \def\c/{{\sf eax$_0$} $<_s$ {\sf 0}}
    \begin{tabular}{lcl}
      {\sf tmp$_{64}$} $\leftarrow$  {\sf sext$_{64}$} {\sf eax$_0$} & \\

      {\sf edx$_0$} $\leftarrow$ {\sf extract$_{32 .. 63}$} {\sf tmp$_{64}$} &
                                                                               $\hookrightarrow$
                                                                             &

{\sf edx$_0$} $\leftarrow$  \c/ ? 0xffffffff : 0 \\

      {\sf eax$_1$} $\leftarrow$  {\sf eax$_0$} $\bm\oplus$ {\sf edx$_0$}
                                                                      & $\hookrightarrow$
      & {\sf eax$_1$} $\leftarrow$ \c/ ? $\neg {\sf eax}_0$ {\sf :} {\sf eax$_0$} \\

      {\sf eax$_2$} $\leftarrow$ {\sf eax$_1$} $\bm-$ {\sf edx$_0$}
      & $\hookrightarrow$
      & {\sf eax$_2$} $\leftarrow$ \c/ ?
      {\sf -eax$_0$} {\sf :} {\sf eax$_0$}
      \\
    \end{tabular}}
    \caption{IR transformations}
    \label{fig:expr-propagation-ir}
  \end{subfigure}
  \caption{Expression propagation}
  \label{fig:expr-propagation}
\end{figure}

\Cref{fig:expr-propagation} shows how the addition of rewrite rules
exposes the intended semantics of a branchless absolute value computation.

\paragraph{Loop normalization (threat \ref{threat:loop})} \label{sec:loop-normalization}
This pass aims at highlighting the relations between the
current iteration of the loop and the variable values. We especially look for
affine relations of the form $a \times x + b$ where $x$ is the loop iteration
counter.  We  indeed found out that tools much prefer to analyze
\cinl{for (int i = 0; i < N; i++) T[i] = C;} instead of
\cinl{for (char *t = T; t < T + N; t++) *t = C;}.
Assembly code, though, is more likely to have the second form.

We thus transform
each self-incrementing (-decrementing) variables of the
form \cinl{v = I; while (...) { ...; v = v + k; }} in order to get code more
amenable to analysis. The transformation is done in (up to) 3 steps:
\begin{enumerate}
\item \textbf{rebasing} replaces the initial value \cinl{I} by 0 and each occurrence of
the variable \cinl{v} by \cinl{I + v};
\item \textbf{rescaling} replaces the increment \cinl{k} by 1 and each
occurrence of the variable \cinl{v} by \cinl{k * v};
\item \textbf{merging} unifies the transformed variables with the loop iteration counter.
\end{enumerate}

For example, in
\cref{fig:motivation_example_tina_result}, the byte-level affine relation between
the counter \cinl{ecx}, lifted as \cinl{__tina_ecx}, and the moving pointer
\cinl{edi}, based at \cinl{__tina_4}, is  \cinl{edi} $\equiv$
\cinl{__tina_4 + 4 * (__tina_3 - ecx)} --- the code is
lifted as \cinl{__tina_4 + (__tina_3 - __tina_ecx)} to take
pointer arithmetic into account (\cinl{__tina_4} is an \cinl{int *}, pointing to 4 bytes long
values in \x86/).

\section{Validation}
\label{sec:validation}

For our translation to be trustworthy, we use a
two-pronged approach:
\begin{enumerate*}[label=\arabic*)]
\item We try to prove the semantic equivalence of the code prior to lifting
 with the lifted C code;
\item If this fails, we rely on intensive random testing (fuzzing) to increase
  the level of trust in the lifted C code.
\end{enumerate*}

\paragraph{Block-based semantic equivalence}
The lifting process of~\cref{sec:lifting} strives to preserve the isomorphism of
the control-flow graphs based on basic blocks between the initial assembly chunk
and its lifted C representation over their DBA IR representation.  This property
allows us to tackle the equivalence proof at basic block level. The proof of
equivalence proceeds as follows:

\begin{enumerate}[label={\bf S\arabic*}.,ref={\bf S\arabic*},wide]
\item \label{seq:isog} We {\em check the isomorphism of the control-flow graphs
    extracted from the two lifted programs}. Since we deal with deterministic
  labeled directed graphs, this check is immediate --- and usually
  succeeds. \approach/ is actually very careful during simplifications and
  recompilation to preserve the control-flow structure (see details below).
  For the isomorphism check, we track the relation between the heads
  of IR basic blocks and the corresponding emitted C code thanks to C labels and
  debug line information.
  If the check succeeds, we go to \ref{seq:block}, otherwise we {\bf [fallback]} on
  fuzzing  --- in practice~(\cref{sec:applicability}), the latter has never happened.
\item \label{seq:block} Once we know the two control-flow graphs are
  isomorphic, we \emph{try to demonstrate the pairwise equivalence of corresponding vertices}.
  This allows to avoid directly dealing with loops. Each pairing of basic blocks is
  translated to logical formulas for which we ask SMT solvers: if  inputs are
  identical, can outputs be different? If all queries are unsatisfiable then
  {\it equivalence is proven} {\bf [success]}, otherwise we use our {\bf [fallback]}.
\end{enumerate}

\paragraph{Taming simplifications}
In order to help the equivalence proof succeed, \approach/ passes were designed to \emph{preserve the
control-flow graph structure} and \emph{to be traceable}.
For the first goal,
simplifications never modify jump instruction, except for trivial dead branch
elimination and the lifter avoids inserting branches with lazy constructions
such as \cinl{&&}, \cinl{||} or ternary operators.
For the second goal, when a
simplification changes the input-output relation of a basic
block, it records the changes w.r.t the old ones and these properties
will be added to the assumptions of \ref{seq:block}. For instance, in
\cref{fig:running_example}, the expression propagation records that \cinl{eax}
holds the value \cinl{0} for the entire chunk. It will then be used during
\ref{seq:block} to prove the equivalence of the loop body where the register no
longer exists in the generated part (\cref{fig:motivation_example_tina_result}).

\medskip
\noindent \textbf{What could go wrong?} While \approach/ uses simplifications
and lifting passes tailored to make the block-based semantic equivalence
algorithm possible, the recompilation step is blind to this requirement and may
therefore threaten it.

The \ref{seq:isog} check may fail if the compiler
  modifies the control flow graph, for example if some elements outside of the
  assembly chunk render a branch dead or a loop unrollable. In
  \cref{fig:motivation_example_tina_result}, since \cinl{sizeof} is
  known at compile time,
  \clang/ \textsf{-O1} unrolls the loop, making the
  isomorphism check fail.

  The \ref{seq:block} query may fail if the compiler moves parts of the
  computation across basic blocks, changing the relation between inputs and
  outputs. It may happen during code motions, like loop-invariant code
  motions. In this case, the graph isomorphism still holds but the relation
  between basic blocks is lost.
  \gcc/ {\sf -funroll-loops} partially unrolls
  (8 times) the loop body in
  \cref{fig:motivation_example_tina_result} leading to a failed equivalence query.

To avoid such problems, we recompile the code {\em without any optimization} (-O0).

Note that SMT checks never time out in our experiments (\cref{sec:xps}),
probably due to the naturally small size of block-based queries. However, we can
imagine that code showing hard-to-reverse behaviors, such as
cryptographic hash functions, could make the \ref{seq:block} query fail.

\paragraph{Trust base}
\label{sec:trustbase}
Validation allows to increase the confidence in the lifting process,
using 3 components as the {\em trust base}: the binary-code lifter,
the compiler and the solver. All are well tested software and the last two
are part of the trust base of (most) modern source-level verification tools
anyway. Furthermore, while we trust the compiler debug information, we argue
that the compilation process itself is not part of the trusted base:
assembly chunks are untouched by it and validation will very likely catch
errors during re-compilation. Besides, further mitigation includes
systematic testing of assembly chunks vs. their IR representation,
and using  multiple compilers and/or solvers.

\newcommand{\pc}[1]{{\tiny #1\%}}
\newcommand{\upto}[1]{{\scriptsize (#1)}}
\newcommand{\ccent}[1]{\multicolumn{1}{c}{#1}}
\newcommand{\dual}[1]{\multicolumn{2}{c}{#1}}
\newcommand{\light}[1]{\arrayrulecolor{lightgrey}#1\arrayrulecolor{darkgrey}}
\newcommand{\white}[1]{\arrayrulecolor{white}#1\arrayrulecolor{darkgrey}}
\begin{table*}[t]
  \caption{Applicability on Debian 8.11 Jessie distribution (\gcc/ 5.4)}
  \centering
 \begin{tabular}{lr@{\hspace{2pt}}rr@{\hspace{2pt}}rr@{\hspace{3pt}}rr@{\hspace{3pt}}rr@{\hspace{3pt}}rr@{\hspace{3pt}}rrr@{\hspace{3pt}}rr@{\hspace{3pt}}rr@{\hspace{3pt}}r}
    & \multicolumn{12}{c}{\x86/} & \multicolumn{7}{c}{\arm/} \\
    \light{\cmidrule(lr){2-13}\cmidrule(lr){14-20}}
     & \dual{\scshape Total} & \dual{\scshape Big 100} & \dual{\alsa/}  & \dual{\ffmpeg/} & \dual{\gmp/} & \dual{\libyuv/} & \ccent{\alsa/}  & \dual{\ffmpeg/} & \dual{\gmp/} & \dual{\libyuv/} \\
    \light{\cmidrule(lr){2-3}\cmidrule(lr){4-5}\cmidrule(lr){6-7}\cmidrule(lr){8-9}\cmidrule(lr){10-11}\cmidrule(lr){12-13}\cmidrule(lr){14-14}\cmidrule(lr){15-16}\cmidrule(lr){17-18}\cmidrule(lr){19-20}}
    Assembly chunks & 3039 &           & 100 &           & 25 &          & 103 &          & 237 &          & 4 &           & \ccent{0}           & 85 &          & 308 &        & 1 &            \\
    \white{\midrule}
    Trivial         & 126 &            & 0 &             & 0 &           & 6 &            & 13 &           & 0 &           & \ccent{--}          & 1  &         & 28  &        & 0 &            \\
    Out-of-scope    & 449 &            & 40 &            & 0 &           & 17  &          & 0  &           & 3 &           & \ccent{--}          & 2  &         & 0   &        & 0 &            \\
    Rejected        & 138 &            & 11 &             & 0 &           & 12  &          & 1 &            & 0 &           & \ccent{--}          & 12 &          & 2 &          & 0 &            \\
    \white{\midrule}
    Relevant        & 2326 & \pc{76}  & 49 & \pc{49}   & 25 & \pc{100} & 68 & \pc{66}  & 223 & \pc{94}  & 1 & \pc{25}  & \ccent{--} & 70 & \pc{82} & 278 & \pc{90}  & 1 & \pc{100}  \\
    Lifted          & 2326 & \pc{\bf 100} & 49 & \pc{\bf 100}  & 25 & \pc{\bf 100} & 68 & \pc{\bf 100} & 223 & \pc{\bf 100} & 1 & \pc{\bf 100} & \ccent{--} & 70 & \pc{\bf 100} & 278 & \pc{\bf 100} & 1 & \pc{\bf 100} \\
    Validated       & 2326 & \pc{\bf 100} & 49 & \pc{\bf 100}  & 25 & \pc{\bf 100} & 68 & \pc{\bf 100} & 223 & \pc{\bf 100} & 1 & \pc{\bf 100} & \ccent{--} & 70 & \pc{\bf 100} & 278 & \pc{\bf 100} & 1 & \pc{\bf 100} \\
    \white{\midrule}
    Average {\scriptsize (Max)} size    & 8 & \upto{341}        & 104 & \upto{341}           & 50 & \upto{70}           & 5 & \upto{10}           & 6 & \upto{31}            & 31 & \upto{31}          & \ccent{--} & 5 & \upto{16}  & 5 & \upto{10}  & 29 & \upto{29}         \\
   Lifting time (s) & 121 & & 98 & & 2 & & 63 & & 2 & & $<$ 1 & & \ccent{--} & $<$ 1 & & 4 & & $<$ 1  \\
   Validation time (s) & 1527 & & 36 & & 17 & & 255 & & 110 & & $<$ 1 & & \ccent{--} & 48 & & 187 & & $<$ 1  \\
  \end{tabular}

  \label{tab:wa-debian}
\end{table*}

\section{Experimental evaluation}
\label{sec:xps}

We evaluate our implementation of \approach/ on 3 research questions:
\begin{enumerate*}[label={\bf RQ\arabic*)},ref={\bf RQ\arabic*}]
\item \label{item:rq1} How applicable is it on assembly chunks found in the wild?
\item \label{item:rq2} How do off-the-shelf  program analyzers behave on lifted
  code?
\item \label{item:rq3}  What  is the impact of each optimization?
\end{enumerate*}

\subsection{Wide applicability {\bf (\ref{item:rq1})}}
\label{sec:applicability}

We run our prototype on {\it all} assembly chunks found in the Linux  Debian
8.11 distribution (for \x86/), i.e.~$\approx 3000$ chunks distributed over
200 packages and 1000 functions.
As chunk distribution is not smooth, we also fix 2 subsets of samples: one with
the 100 biggest chunks, and another with all chunks from 4 key major projects
exploiting low-level optimizations: \gmp/, \ffmpeg/, \alsa/ and \libyuv/.
\cref{tab:wa-debian} sums up the results of lifting with \approach/.
\begin{table}[!htbp]
  \caption{Applicability by compiler (\x86/)}
  \centering
  \begin{tabular}{lclclcl}
    & \dual{\gcc/ 5.4} & \dual{\gcc/ 4.7} & \dual{\clang/ 3.8} \\
    \light{\cmidrule(lr){2-3}\cmidrule(lr){4-5}\cmidrule(lr){6-7}}
    Assembly chunks & 3039 & % \pc{100}
                                  & 2955 & % \pc{97}
                    & 2852 & % \pc{94}
    \\
    \white{\midrule}
    Relevant        & 2326 & \pc{76}  & 2326  & \pc{78} % \pc{76}
                    & 1970 & \pc{69} % \pc{65}
    \\
    Lifted          & 2326 & \pc{\bf 100} & 2326  & \pc{\bf 100}  & 1970 & \pc{\bf 100} \\
    Validated       & 2326 & \pc{\bf 100} & 2326  & \pc{\bf 100}  & 1970 & \pc{\bf 100} \\
  \end{tabular}
  \label{tab:wa-compilers}
\end{table}

We exclude trivial (empty or unused), out-of-scope and rejected chunks.
{\it Out-of-scope chunks} include those with floating point operations, OS-level
hardware instructions  or hardware-based crypto-primitives, like AES. {\it Rejected
chunks} are those deemed unsafe because they do not respect their interface.
Yet, we activate options in our tool to specifically
regard accessing flags, \cinl{xmm} registers or memory as safe --  allowing to
consider 150 extra chunks as relevant, notably in \ffmpeg/. The statistics of
\cref{tab:wa-debian} report on the tool's behavior with these settings.

On in-scope chunks, \approach/ {\it performs extremely well}, with {$\bf 100\%$}
chunks lifted {\em and} fully validated (no resort to testing) --- this amounts
to 76\% of all chunks found --- {\it for a negligible cost} (0.7s per chunk on
average). The biggest 100 chunks are a little less successful as they have a fair
amount of (unhandled) floating-point instructions. \approach/ works equally well
on major projects for \arm/ or \x86/, and with \gcc/ or \clang/ on \x86/
(\cref{tab:wa-compilers}), confirming its genericity.

\subsection{Adequacy to formal verification tools {\bf (\ref{item:rq2}, \ref{item:rq3})}}
\label{sec:verification}

We select 3 tools representing popular formal techniques
{\it currently used in the industry}:  KLEE
\cite{DBLP:conf/osdi/CadarDE08} for symbolic execution
\cite{DBLP:journals/cacm/CadarS13} (bug finding),  and
Frama-C \cite{DBLP:journals/fac/KirchnerKPSY15} with its EVA plug-in \cite{DBLP:phd/hal/Buhler17} for abstract
interpretation \cite{DBLP:conf/popl/CousotC77,DBLP:conf/csl/CousotC14} (runtime error verification) and WP
plug-in \cite{WPmanual} for deductive verification \cite{Floyd67, DBLP:journals/cacm/Hoare69} (functional correctness).

Experiments on both symbolic execution and abstract interpretation use
58 functions (out of 366) from the 4 key projects  in \cref{sec:applicability}, selected due to the ease of
automatically generating the initial contexts for both analyses.  For all 3
tools, we also report the observed differences using a {\em basic} lifter and
different optimization levels: {\em O1} (high-level predicate recovery), {\em
  O2} (O1 + register unpacking), {\em O3} (O2 + expression propagation) and {\em
  O4} (O3 + loop normalization).  Note that \textbf{\textit{O4} is \approach/}.

\begin{table}[!htbp]
  \caption{Impact of \approach/ \& lifting strategies on KLEE}

  \resizebox{\linewidth}{!}{
  \begin{tabular}{p{3.1cm}cccccc}
    &  \multicolumn{6}{c}{\sc Lifting} \\
        \light{\cmidrule(lr){2-7}}
     &  {\sc None } & {\sc Basic } & O1 & O2 & O3 & O4 \\
    \light{\cmidrule(lr){2-2}\cmidrule(lr){3-3}\cmidrule(lr){4-4}\cmidrule(lr){5-5}\cmidrule(lr){6-6}\cmidrule(lr){7-7}}
    Functions analyzed w/o blocking & 3 & 58 & 58 & 58 & 58 & 58 \\
    \light{\cmidrule(lr){1-7}}
    Functions 100\% covered & {\bad} & 25 & 25 & 25 & 25 & 25 \\
    Aggregate time & \na/ & 115s & 115s & 110s & 103s & 105s \\
    \light{\cmidrule(lr){1-7}}
    \# paths (all functions) & 1.4M & 1.5M & 1.8M & 4.6M & 6.6M & 6.6M \\
  \end{tabular}}

  \label{tab:klee}
\end{table}

\paragraph{Symbolic execution}
\label{sec:se}
We perform our experiments with KLEE~\cite{DBLP:conf/osdi/CadarDE08} which at
present does not handle inline assembly chunks and stops upon meeting one --
except for a very few simple cases such as assembly-level rotations.  This fact
can sometimes prevent the adoption of symbolic execution~\cite{AsmKlee18}.

\Cref{tab:klee} summarizes our findings. (\ref{item:rq2}) First, KLEE alone can
analyze only few functions (3/58) as (almost any block of) assembly stops the
analysis, and none of them is fully path-covered. Adding lifting allows to {\it
  analyze all considered functions} (58/58), to completely path-cover 43\% of
them (25/58) and to explore significantly more paths within the same analysis
 budget ($\times 4.7$).

The lifting strategy (\ref{item:rq3})  does
not impact the functions that KLEE can  fully cover, but \approach/ optimizations considerably
speed up code exploration, enabling to cover
significantly more paths ($\times 4$) than basic lifting
in the same amount of time. This is explained by \approach/-produced code being higher-level, with
fewer instructions and local variables,
thereby accelerating SMT-solving.
Note that control-flow structure, and thus total number of paths, does not change.
Moreover, each optimization step brings some degree of improvement. The major improvement gaps here are brought by
register unpacking (O2) and expression  propagation (O3). As expected, loop normalization (O4) has no impact as
symbolic execution simply unrolls loops. Additional experiments
\begin{long-paper}
  (Supplementary material, \cref{tab:klee-ext})\phantom{ }
\end{long-paper}
demonstrates that high-level recovery (O1) has also a substantial impact on the analysis
(removing it leads to 5.4M explored paths, vs 6.6M in full \approach/).

\paragraph{Abstract interpretation}
\label{sec:abstr-interpr}
We use the Frama-C EVA \cite{DBLP:phd/hal/Buhler17} plug-in.
 Frama-C has limited support for inline assembly based on
interfaces, translating them into logical {\sf assigns}
annotations for modified variables -- safely interpreted
in EVA (and WP) as non-deterministic assignments.

\newcommand{\mr}[1]{\multirow{2}{*}{#1}}

\Cref{tab:eva} sums up the results for \textbf{\ref{item:rq2}}. Lifting the assembly code with \approach/ almost always {\it reduces the
number of alarms in the common C code} (23/27). This follows from the better
precision of the analysis since modified variables in the lifted code are now
accessible. In half the cases (11/20), we observe a {\it precision gain on function
return values}. Most functions (31/34) with return values or
initial C alarms show such improvements.

\newcommand{\on}[1]{\tiny / #1}
\begin{table}[!htbp]
  \caption{Impact of \approach/ on EVA}
  \centering
  \begin{minipage}{1.0\linewidth}
    \begin{tabular}{lrrrrr@{\hspace{2pt}}l}
      \multicolumn{1}{c}{Function with}                   & \multicolumn{1}{c}{\alsa/} & \multicolumn{1}{c}{\ffmpeg/} & \multicolumn{1}{c}{\gmp/} & \multicolumn{1}{c}{\libyuv/} & \multicolumn{2}{c}{\sc Total}
      \\
      \light{\cmidrule(lr){1-1}\cmidrule(lr){2-2}\cmidrule(lr){3-3}\cmidrule(lr){4-4}\cmidrule(lr){5-5}\cmidrule(lr){6-7}
      }
      \scriptsize Returns (non \cinl{void}) & \scriptsize 0 & \scriptsize 9 & \scriptsize 10 & \scriptsize 1 & \scriptsize 20 \\
      \scriptsize Better return values
                                                          & \scriptsize --        & \scriptsize 9
                                                                              & \scriptsize 1  & \scriptsize 1      & \scriptsize 11  & \tiny 55\%
      \\

      \white{\midrule}
      \scriptsize Initial C alarms & \scriptsize 2 & \scriptsize 8 & \scriptsize 16 & \scriptsize 1 & \scriptsize 27 \\
      \scriptsize Alarm reduction in C   & \scriptsize 2         & \scriptsize 8            & \scriptsize 12    & \scriptsize 1      & \scriptsize 23  & \tiny 85\%
      \\

      \white{\midrule}
      {\scriptsize New memory alarms} % \asm/
                                                          & \scriptsize 12 & \scriptsize 2            & \scriptsize  3       & \scriptsize 0  & \scriptsize  17 & \tiny 26\%
      \\
      \white{\midrule}
      \multicolumn{1}{r}{Positive impact}
                                                          & 14 & 17           & 13       & 1  & 45 & \tiny 77\%
    \end{tabular}

  \end{minipage}

  \label{tab:eva}
\end{table}

The lifted C code  also contains {\it new alarms} (17/58)  which we could not detect
before and should be taken into account (usually out-of-bounds or other memory
accesses). We also found some {\it possibly buggy behaviors} (\cref{sec:xp-findings}).

For short, we observe positive impact from \approach/ w.r.t. non-lifted code
on $77\%$ (45/58) of the
functions (more precision, reducing alarms from over-ap\-pro\-xi\-ma\-tions
of inline assembly, or new memory alarms in lifted code) .

\begin{table}[htbp]
  \caption{Impact of lifting strategies on EVA}

  \begin{tabularx}{\linewidth}{Xcccccc}
     & \multicolumn{6}{c}{\sc Lifting} \\
    \light{\cmidrule(lr){2-7}}
  \# Functions & \sc None & \sc Basic & O1 & O2 & O3 & O4 \\
    \light{\cmidrule(lr){1-1}\cmidrule(lr){2-2}\cmidrule(lr){3-3}\cmidrule(lr){4-4}\cmidrule(lr){5-5}\cmidrule(lr){6-6}\cmidrule(lr){7-7}}
    without any alarms        & \bad & 12 & 12 & 14 & 14 & {\bf 19} \\
    with \asm/ memory\\
    \multicolumn{1}{r}{alarms}  & \na/  & 29 & 29 & 29 & 21 & 17 \\
    \multicolumn{1}{r}{\em errors} & \bad  & 1 & 1 & 1 & 2 & 2 \\
    \white{\midrule}
    emitted C alarms    & 231 & 184 & 184 & 177 & 177 & 177 \\
    emitted \asm/ alarms & \na/ & 316 & 244 & 199 & 165 & 128 \\
    total  alarms & 231 & 500 & 428 & 376 & 342 & 305
  \end{tabularx}
  \label{tab:eva-cmp}
\end{table}

\Cref{tab:eva-cmp} additionally shows the impact of the lifting strategy (\ref{item:rq3}).
Compared with basic lifting, each additional optimization increases the quality
of the lifted code (fewer ASM and total alarms) and the precision of the analysis (more
functions without alarms, fewer memory alarms, more errors) -- including loop normalization which allows finer approximations of loop fixpoints ({\it widening}). \approach/ (O4)
thus significantly improves all these aspects. Moreover, the produced alarms
are more precise: possible buffer overflows
(such as a \ffmpeg/ -1 index access\begin{long-paper}
  \phantom{ }-- see Appendix~\ref{sec:example-ffmpeg}
\end{long-paper}) are now recognized as errors and not mere alarms.
Additional experiments
\begin{long-paper}
  (Supplementary material, \cref{tab:klee-ext})\phantom{ }
\end{long-paper}
demonstrates that
removing any of the optimization steps leads us quite far from the whole chain result.

\begin{table}[!htbp]
  \caption{Impact of \approach/ \& lifting strategies on WP}
   \centering
  \begin{tabular}{lcccccc}
    & \multicolumn{6}{c}{\sc Lifting} \\
        \light{\cmidrule(rl){2-7}}
    \multicolumn{1}{c}{\sc Function} & {\sc None} & {\sc Basic } & {\sc O1} & {\sc O2} & {\sc O3} & {\sc O4}
    \\
    \light{\cmidrule(rl){1-1}\cmidrule(rl){2-2}\cmidrule(rl){3-3}\cmidrule(rl){4-4}\cmidrule(rl){5-5}\cmidrule(rl){6-6}\cmidrule(rl){7-7}}
    {\sf saturated\_sub} & \bad & \good & \good & \good & \good & \good
    \\
    {\sf saturated\_add} & \bad  & \bad & \good & \good & \good & \good
    \\
    {\sf log2}           & \bad & \bad & \bad & \bad & \good & \good
    \\
    {\sf mid\_pred}      & \bad & \bad & \good & \good & \good & \good
    \\

    {\sf strcmpeq}       & \bad & \bad & \bad & \bad & \good & \good
    \\
    {\sf strnlen}        & \bad & \bad & \bad & \bad & \good & \good
    \\

    {\sf memset}         & \bad & \bad & \bad & \bad & \good & \good
    \\
    {\sf count}          & \bad & \bad & \bad & \bad & \good & \good
    \\
    {\sf max\_element}   & \bad & \bad & \good & \good & \good & \good
    \\
    {\sf cmp\_array}     & \bad & \bad & \bad & \bad & \good & \good
    \\

    {\sf sum\_array}     & \bad & \bad & \bad & \bad & \good & \good
    \\

    {\sf SumSquareError} & \bad & \bad & \bad & \bad & \good & \good
\\
  \end{tabular}

  \label{table-wp}
\end{table}

\paragraph{Weakest precondition calculus}
We use the deductive verification Frama-C plug-in  WP \cite{DBLP:conf/nfm/CarvalhoSPT14,WPmanual}.
We take 12 assembly-optimized functions\begin{long-paper}
  \phantom{ }(see details in Supplementary, \cref{tab:wp-descr})
\end{long-paper}: 6 excerpts
from \ffmpeg/, \gmp/, \libyuv/, \libgcrypt/ and \udpcast/, 2 others adapted from
optimized assembly snippets and 4 translated examples from ACSL by example
\cite{acslbyexample}.
Functional specifications and loop invariants are manually inserted before
verification, as usual for WP-based methods -- we do not insert any other
annotation.
Moreover, recall that without lifting, assembly chunks are correctly
over-approximated by non-deterministic assignments to the modified C
variables.

\cref{table-wp} details our results.
The unlifted code does not require invariants (no C-level loops), while lifted
codes all require identical invariants as they share the same control-flow structure.
A quick glance at \cref{table-wp} shows that (\ref{item:rq2}) while WP without
lifting \emph{never} succeeds and basic lifting is far from enough (1/12),
\approach/ does allow {\it to prove the functional correctness of all functions}
(12/12).
The simple over-approximations of assembly chunks provided by Frama-C without lifting are not sufficient to prove
properties as strong as functional correctness.

Regarding optimization steps (\ref{item:rq3}), it turns out that
 loop normalization (O4) has no direct impact since the user must provide manual
loop invariants. On the other hand, all other steps are complementary (\cref{table-wp}) and crucial:
removing only one of them yields at best a 6/12 success rate\begin{long-paper}
  \phantom{ }(Supplementary material, \cref{tab:klee-ext})
\end{long-paper}.

\subsection{Conclusion}
\label{sec:xp-conclusion}

Experiments
show that
our code lifting method is {\it highly practical} (100\% Debian 8.11 in-scope blocks
  are lifted and validated), that it has a {\it positive and significant
  impact  on all 3 formal verification tools
  considered} -- allowing them to effectively handle code with inline assembly,
and, finally, that full \approach/ (O4) is needed to facilitate further code
analyses -- as less refined lifting yields poorer analyses.

Interestingly, all analyses do not behave the same w.r.t the optimization chain:
symbolic execution mostly takes advantage of register unpacking and expression simplifications,
abstract interpretation is sensitive to the 4 optimization steps and weakest precondition calculus
strongly requires all of them  but loop normalization -- which is already granted by user-supplied loop invariants.

\subsection{Epilogue: post-analysis considerations}
\label{sec:xp-findings}

We found {\bf 567} compliance issues during our
experiments.
Most have no impact with {\it current} compilers but may induce
bugs  out of compiler changes, maintenance or code reuse.

While evaluating verifiability, we ran into \emph{6} potential buffer overflows
hidden in assembly chunks.  For example, a \ffmpeg/ function accesses index
\cinl{-1} of its input buffer -- this is actually reported in the comments\begin{long-paper}
  \phantom{ }(see details in Appendix~\ref{sec:example-ffmpeg})
\end{long-paper}.
All errors  initially reported by Frama-C EVA  were also reproduced with KLEE. After
determining and adding relevant logical preconditions, we were able to show the absence
of runtime errors in the reported ``corrected'' functions.
Besides, we were able to prove (with Frama-C WP) the functional correctness of 6
functions from the Debian distribution code base, including
\cinl{SumSquareError} (24 assembly instructions).

\section{Discussion}
\label{sec:discussion}

\subsection{Threats to validity}

\paragraph{Benchmark representativeness}
The considered code base is quantitatively and qualitatively
representative of the use of inline assembly:  it is extensive and comprises
highly popular and respected projects.  We mainly experiment on \gcc/ and \x86/,
but our experiments on \arm/ and \clang/ show our results also hold in these
settings.
Still, we obviously miss closed-source software and code which relies on
Microsoft's C compiler (different assembly syntax). Yet, there is no reason to
believe it would behave differently.

\paragraph{Verification methods} We consider three of the most popular
verification techniques (symbolic execution, abstract interpretation,
deductive verification), representative of the major classes of analysis, both
in terms of goal (bug finding, runtime error checking and proving functional
correctness)
and underlying core technologies (domain propagation, constraint
solving \& path exploration, first-order reasoning).
Also, we rely on well-established verification tools, each applied in several
successful industrial case studies.
Thus, we reckon that our experiments support our claim regarding the
general verifiability of the codes \approach/ produces.

\subsection{Limitations}

Our lifting has two main limitations: hardware-related instructions
and floating-point operations.

Since we aim to lift assembly chunks back to C, the support of hardware
related instructions cannot be achieved outside of modeling hardware in C as
well --- for example, neither DBA IR nor C can make direct reference to hardware
interrupts. Here we probably cannot do better than having two (approximated) C
models of hardware instructions, one for over- and one for
under-approximations. While not necessarily that difficult for reasonable
analysis precision, this is clearly a manpower-intensive task.

The
float limitation is primarily  due to the lack of support in \binsec/.
Adding such support is also manpower-intensive, but not that hard.
Yet, the real issue is that efficient reasoning over floats is still ongoing
scientific work in both program analysis and automated solvers (e.g., theory
support is new in SMT-LIB \cite{BarFT-SMTLIB}, only 2 solvers in the relevant
category of SMT-COMP 2018).  %
As such, it threatens our validation part, and most program analyzers would not be able
to  correctly handle these lifted floats anyway.
\emph{Despite these limits, we still lift and validate  76\% of assembly chunks of a standard Linux distribution}.

Finally, our technique is amenable, to a
certain extent, to standalone assembly code or even binary code decompilation.
However, this case can quickly
deteriorate to the usual difficult problem of lifting an arbitrary program.
Especially, dynamic jumps or large-size complicated CFG would probably
yield serious issues.

\section{Related Work}
\label{sec:reference}

Though some prior work has addressed code lifting for verification,
it is worth noting that {\it verifiability} has never been explicitly addressed so far.
We hereafter review approaches (partly) related to our method.

\paragraph{Assembly code lifting and verification}
Maus \cite{Maus2008, DBLP:phd/dnb/Maus11} proposes a generic method  simulating the behavior of assembly instructions in a virtual machine written in
C. This work was used by the Verisoft project to verify the code of an
hypervisor consisting of mixed low-level code. Maus' technique relies on {\sf
  VCC} \cite{Cohen2009} to write and prove verification conditions regarding the
state of its machine. While we strive to produced high-level code, Maus'
virtual code contains all low-level code details, including flags.

Further work by Schmaltz and Shadrin \cite{Schmaltz2012} aims (only) at proving the ABI
compliance of the assembly chunks. This method is however restricted to MASM and
the Windows operating system. \approach/, here applied to  \gcc/
inline assembly, is independent of the assembly dialect by leveraging
binary level analyzers and is applicable to a wider range of architectures.

Fehnker et al. \cite{Fehnker08} tackle the analysis of inline assembly for ARM
architecture, using a model-checking based syntactic analysis to integrate \corcpp/
analyses with inline assembly. This solution is however limited by its purely
syntactic basis: first, it is restricted to one single inline assembly dialect;
second it loses the soundness properties we target. Losing soundness may be an
appropriate practical trade-off but not when targeting sound formal
analyses.

Corteggiani et al. \cite{DBLP:conf/uss/CorteggianiCF18} also use code lifting
within their framework. However, their end goal is to perform dynamic symbolic
analyses on the produced lifted code. \Cref{sec:verification} shows that such
very targeted lifting may not be enough for other formal analyses. Moreover,
correctness of the translation is not addressed.

Myreen et al. \cite{4689183} targets verification of pure assembly code. The
translation corresponds to our basic lifter, yet the approach proves the initial
lifted IR is semantically equivalent to a very detailed ISA model. This paper
then targets verification at the level of assembly code but requires code
annotations and interactive proving.  Our proposal targets
the lifting of inline assembly within C for (general) verification purposes, is
geared at ensuring the verifiability of the produced code, and its validation
establishes the correctness of the IR transformations producing the final
extracted C code.

\paragraph{Decompilation}
Decompilation
\cite{DBLP-conf/icsm/CifuentesSF98,DBLP-journals/jpl/Cifuentes96,DBLP-journals/spe/CifuentesG95}
tackles the challenge of recovering the original
source code (or a similar one) from an executable. This goal is very difficult
and requires hard work to find back the information lost during compilation \cite{Chang2006}.
 Despite significant recent progress
\cite{DBLP:conf/uss/BrumleyLSW13}, decompilation  remains an open
challenge.  Still, it is used to enhance program understanding, e.g.,
during reverse engineering. As such, correctness is not the main concern
--- for example it does not
always need produce compilable source~code.

Soundness is addressed by two recent works. Schulte et al. \cite{SchuRuNoCiLo18}
use search-based techniques to generate source-code producing byte-equivalent
binaries to the original executable. This technique, when it succeeds, ensures
soundness by design but it is only applied to small examples, with limited success.
Brumley et al. \cite{DBLP:conf/uss/BrumleyLSW13} on the other hand use testing to increase
trust~in~their~lifted~code.

We do draw inspiration from some decompilation techniques for type
reconstruction \cite{Lee2011TIEPR, Robbins:2016:MMS:2837614.2837633}. Even
though we do not construct types that are not derived from inputs, it helps in
strengthening our type system.

Recovering the instructions and CFG of the code under analysis is a big
challenge in decompilation
\cite{DBLP:conf/uss/AndriesseCVSB16,DBLP:conf/issta/MengM16}, especially for adversarial codes like
malware.  The regularity
and patterns of managed codes allow a very good recovery in practice
\cite{Lee2011TIEPR} by unsound methods, yet without any guarantee. Inline assembly chunks have more limited behaviors (clear control-flow,
no dynamic jumps) and the fact that we control compilation makes it a non-issue
for us.

\paragraph{Binary-level program analysis}
For more than a decade now, the program analysis community has spent significant
efforts on binary-level codes \cite{DBLP:journals/toplas/BalakrishnanR10},
either to analyze source-less programs (malware, COTS) or to check the code that
is really running. The efforts have mainly been concerned with safe high-level
abstraction recovery
\cite{DBLP-conf/vmcai/BardinHV11,DBLP-conf/vmcai/KinderK12,DBLP-conf/emsoft/ReinbacherB11,Djoudi2016,DBLP-conf/wcre/SeppMS11}
and invariant computation.

Several generic binary lifters have been produced
\cite{Brumley2011,DBLP:conf/cav/BardinHLLTV11,DBLP-conf/kbse/KimFJJOLC17},
reducing complex ISAs to a small set of semantically well-defined
primitives. Though well tested \cite{DBLP-conf/kbse/KimFJJOLC17}, more trust could
be achieved if lifters were automatically derived from something
akin to ARM's formal specifications \cite{DBLP-conf/fmcad/Reid16}.

\paragraph{Mixed code problems}
Morrisett et al. \cite{DBLP:conf/popl/MorrisettWCG98} have proposed Typed
Assembly Language to ensure memory and control flow integrity in low-level
assembly.  Patterson et al. \cite{Patterson:2017:FRM:3062341.3062347}
have exploited the idea to mix low-level code with functional languages. We
borrow some elements to propagate types between C and inline assembly.

\paragraph{Translation validation and code equivalence}
In order to achieve safe lifting, we use translation
validation  \cite{DBLP:conf/pldi/SewellMK13,DBLP:conf/pldi/Necula00,
  DBLP:journals/sttt/Rival04}, a technique also used in CompCert register
allocation \cite{DBLP:conf/esop/BlazyRA10}.
Our formal needs thus rely on well-established and tested tools (here SMT
solvers), usable as blackboxes, instead of a full formal proof of the whole
lifting chain.

Program equivalence checking is considered a challenging verification task. Dedicated
approaches start to emerge, like relational
weakest precondition calculus \cite{DBLP:conf/popl/Benton04} (for proof) or
relational symbolic execution
   \cite{DBLP:conf/icse/PalikarevaKC16}  (for bug finding).

\section{Conclusion}

We have presented \approach/, a method enabling the  analysis of \corcpp/
code mixed with inline
assembly, by lifting the assembly chunks to equivalent C code. This method is the first to generate
well-structured C code amenable to formal analysis through a dedicated principled succession of transformations geared at {\it improving the verifiability of the produced code}.
To boot, translation validation {\it builds trust} into the lifting process.  Thorough experiments on real-world code
show that \approach/ is widely applicable (100\% of
in-scope chunks from Linux Debian Jessie 8.11 are validated)
and that its semantic transformations positively (and significantly)
impact  popular verification techniques.

\bibliography{biblio}

\begin{long-paper}
  \clearpage
  \begin{appendices}

\section{Motivating example}
\label{sec:example-ext}

In \cref{sec:motivation}, \cref{fig:motivation_example_tina_result}
shows the code produced by \approach/. Both Frama-C plugins EVA and WP  were
able to prove the absence of buffer overflow.

The \textsc{Basic} lifter on the code from
\cref{fig:motivation_example_asm_part} produces the code shown in
\cref{fig:motivation_example_basic}. Even if the code remains small, the absence of
simplifications adds a lot of unwanted complexity.  Here, \cinl{__tina_df} is
not propagated, thus the \cinl{__tina_edi} pointer is computed by a (extraneous)
cast following bitwise operations.  These operations actually destroys the
analyzer's abstractions.  On this very example, none of the Frama-C plugins was
able to
prove the absence of error.

\begin{figure}[htbp]
  \centering
  \includegraphics{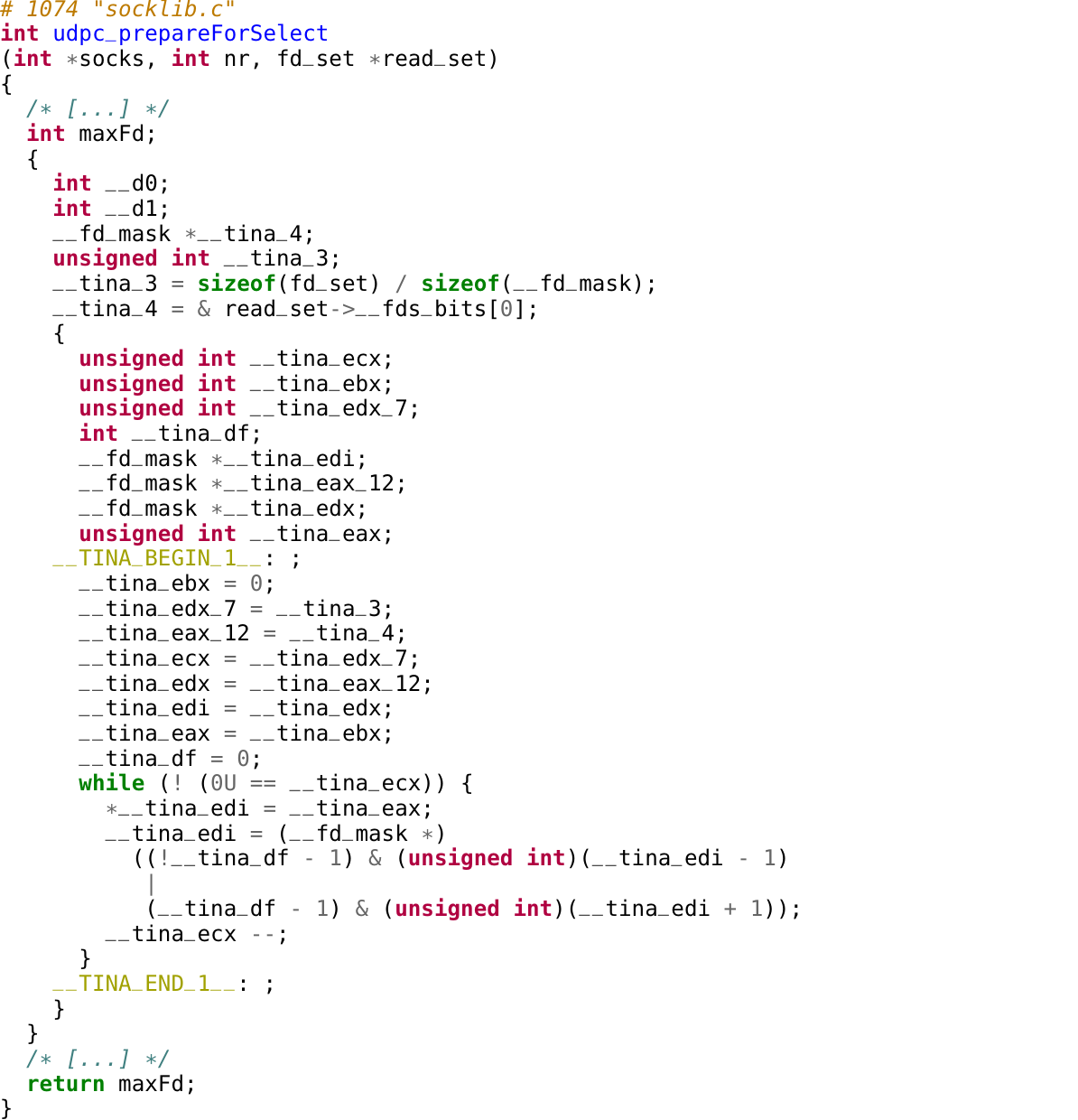}
  \caption{C code generated by \textsc{Basic} lifter}
  \label{fig:motivation_example_basic}
\end{figure}

\newpage
\section{Additional experiments: complementarity of optimization steps}
\label{sec:xp-ext}

\Cref{tab:klee,tab:eva-cmp,table-wp} demonstrated the positive impact of
stacking optimizations (levels $O1$ to $O4$) on common analyzers -- recall that \approach/ is $O4$.
What happens, now, if we remove any of the 3 initial optimizations from \approach/?
Here, we extend previous experiments with a new set of optimization levels:
\begin{itemize}
\item $\overline{O1}$ : $O4$ - high-level predicate recovery;

\item $\overline{O2}$ : $O4$ - register unpacking;

\item $\overline{O3}$ : $O4$ - expression propagation.
\end{itemize}
Note that $\overline{O4}$ (i.e., $O4$ - loop normalization) is actually $O3$, hence this case is not discussed.

\paragraph{Symbolic execution}
\Cref{tab:klee-ext} extends the results of \cref{tab:klee}. We can observe that
deactivating high-level predicate recovery or expression propagation leads to a
drop in the number of paths explored equivalent to what their activation had
gained.

\begin{table*}[htbp]
  \caption{Impact of lifting strategies on KLEE (extended)}
  \centering
  \begin{tabular}{p{3.1cm}ccccccccc}
    &  \multicolumn{9}{c}{\sc Lifting} \\
        \light{\cmidrule(lr){2-10}}
    &  {\sc None } & {\sc Basic } & $O1$ & $O2$ & $O3$ & $O4$ & $\overline{O1}$
    & $\overline{O2}$ & $\overline{O3}$
    \\
    \light{\cmidrule(lr){2-2}\cmidrule(lr){3-3}\cmidrule(lr){4-4}\cmidrule(lr){5-5}\cmidrule(lr){6-6}\cmidrule(lr){7-7}\cmidrule(lr){8-8}\cmidrule(lr){9-9}\cmidrule(lr){10-10}}
    Functions 100\% covered & \bad & 25 & 25 & 25 & 25 & 25 & 25 & 25 & 25 \\
    Aggregate time & \na/ & 115s & 115s & 110s & 103s & 105s & 105s & 105s & 110s \\
    \light{\cmidrule(lr){1-10}}
    \# paths (all functions) & 1.41M & 1.50M & 1.83M & 4.59M & 6.64M & 6.62M & 5.42M & 4.57M & 4.59M \\
                     &       & \tiny +6\% & \tiny +22\% & \tiny +150\% & \tiny +45\% & \tiny $\thicksim$ & \tiny -19\% & \tiny -30\% & \tiny -30\%
    \\
  \end{tabular}

  \label{tab:klee-ext}
\end{table*}

\paragraph{Abstract interpretation}
\Cref{tab:eva-cmp-ext} extends the results of \cref{tab:eva-cmp}. High-level
predicate recovery is very important for abstract interpretation as common
domains do not accurately handle flag computations.  Any imprecision occurring at
a loop exit point will generate a lot of alarms in the loop body and
this is what we observe here.  Register unpacking, in the other hand, has less
impact in our experiments because imprecision on data do not affect the analysis
as much as imprecision on the control flow.

\begin{table*}[htbp]
  \caption{Impact of lifting strategies on EVA (extended)}
  \centering
  \begin{tabular}{lccccccccc}
     & \multicolumn{9}{c}{\sc Lifting} \\
    \light{\cmidrule(lr){2-10}}
  \# Functions & \sc None & \sc Basic & $O1$ & $O2$ & $O3$ & $O4$ & $\overline{O1}$ & $\overline{O2}$ & $\overline{O3}$ \\
    \light{\cmidrule(lr){1-1}\cmidrule(lr){2-2}\cmidrule(lr){3-3}\cmidrule(lr){4-4}\cmidrule(lr){5-5}\cmidrule(lr){6-6}\cmidrule(lr){7-7}\cmidrule(lr){8-8}\cmidrule(lr){9-9}\cmidrule(lr){10-10}}
    without any alarms        & \bad & 12 & 12 & 14 & 14 & {\bf 19} & 14 & 14 & 14 \\
    with \asm/ memory\\
    \multicolumn{1}{r}{alarms}  & \na/  & 29 & 29 & 29 & 21 & 17 & 21 & 21 & 29 \\
    \multicolumn{1}{r}{\em errors} & \bad  & 1 & 1 & 1 & 2 & 2 & 2 & 2 & 1 \\
    \white{\midrule}
    Aggregate of \#\\
    \light{\cmidrule(lr){1-1}}
    emitted C alarms & 231 & 184 & 184 & 177 & 177 & 177 & 177 & 184 & 177 \\
    emitted \asm/ alarms & \na/ & 316 & 244 & 199 & 165 & 128 & 253 & 171 & 199 \\
  \end{tabular}

  \label{tab:eva-cmp-ext}
\end{table*}

\begin{table*}[!htbp]
  \caption{Functions under test for WP}
  \centering
  \small
  \begin{tabularx}{\linewidth}{lccXl}
        {\sc Function} & {\sc \# Inst.} & {\sc \# Inv.} & {\sc Description} & {\sc Origin}\\ \light{\cmidrule(rl){1-1}\cmidrule(rl){2-2}\cmidrule(rl){3-3}\cmidrule(rl){4-4}\cmidrule(rl){5-5}}
     {\sf saturated\_sub} & 2 & 0 & Maximum between 0 and big integer subtraction
                              & \ffmpeg/, \gmp/
    \\
     {\sf saturated\_add} & 2 & 0 & Minimum between {\sf MAX\_UINT} and big integer
                             addition
                                          & \ffmpeg/, \gmp/
    \\
    {\sf log2} & 1 & 5 & Biggest power of 2 of an integer & \libgcrypt/
    \\
    {\sf mid\_pred} & 7 & 0 & Median of 3 inputs & \ffmpeg/
    \\
    \white{\midrule}

    {\sf strcmpeq}  & 9 & 6  & String equality testing   &
      \href{http://www.alfredklomp.com/programming/sse-strings/}{\asm/ snippet} \\
    {\sf strnlen} & 16 & 6 & String length (or buffer length if no \verb?'\0'?)  &
      \href{https://www.strchr.com/optimized_strlen_function}{\asm/ snippet} \\
      \white{\midrule}
    {\sf memset} & 9 & 5 & Set array contents to input & \udpcast/
    \\
    {\sf count} & 8 & 4 & Count occurrences of inputs in array & ACSL by example \cite{acslbyexample}
    \\

    {\sf max\_element} & 10 & 7 & First index of max element of the array & ACSL
                                                                            by example\\
    {\sf cmp\_array} & 10 & 6 & Array equality testing (\simd/) & ACSL by example \\
    {\sf sum\_array} & 20 & 7 & Sum of array elements (\simd/) & ACSL by example \\
    {\sf SumSquareError} & 24 & 69 & Sum of square differences between two arrays (\simd/) & \libyuv/
  \end{tabularx}
  \label{tab:wp-descr}
\end{table*}

\begin{table}[htbp]
  \caption{Impact of lifting strategies on WP (extended).}
  \centering
  \resizebox{\linewidth}{!}{
  \begin{tabular}{lccccccccc}
    & \multicolumn{9}{c}{\sc Lifting} \\
        \light{\cmidrule(rl){2-10}}
    \multicolumn{1}{c}{\sc Function} & {\sc None} & {\sc Basic } & $O1$ & $O2$ & $O3$ & $O4$ & {$\overline{O1}$} & {$\overline{O2}$} & {$\overline{O3}$}
    \\
    \light{\cmidrule(rl){1-1}\cmidrule(rl){2-2}\cmidrule(rl){3-3}\cmidrule(rl){4-4}\cmidrule(rl){5-5}\cmidrule(rl){6-6}\cmidrule(rl){7-7}\cmidrule(rl){8-8}\cmidrule(rl){9-9}\cmidrule(rl){10-10}}
    {\sf saturated\_sub} & \bad & \good & \good & \good & \good & \good & \good & \good & \good
    \\
    {\sf saturated\_add} & \bad  & \bad & \good & \good & \good & \good & \good & \good & \good
    \\
    {\sf log2}           & \bad & \bad & \bad & \bad & \good & \good & \good & \good & \bad
    \\
    {\sf mid\_pred}      & \bad & \bad & \good & \good & \good & \good & \bad & \good & \good
    \\

    {\sf strcmpeq}       & \bad & \bad & \bad & \bad & \good & \good & \good & \bad & \bad
    \\
    {\sf strnlen}        & \bad & \bad & \bad & \bad & \good & \good & \bad & \bad & \bad
    \\

    {\sf memset}         & \bad & \bad & \bad & \bad & \good & \good & \good & \bad & \bad
    \\
    {\sf count}          & \bad & \bad & \bad & \bad & \good & \good & \bad & \good & \bad
    \\
    {\sf max\_element}   & \bad & \bad & \good & \good & \good & \good & \bad & \good & \good
    \\
    {\sf cmp\_array}     & \bad & \bad & \bad & \bad & \good & \good & \bad & \bad & \bad
    \\

    {\sf sum\_array}     & \bad & \bad & \bad & \bad & \good & \good & \bad & \bad & \bad
    \\

    {\sf SumSquareError} & \bad & \bad & \bad & \bad & \good & \good & \bad & \bad & \bad
  \end{tabular}}

  \label{table-wp-ext}
\end{table}

\paragraph{Weakest precondition calculus}
First, \cref{tab:wp-descr} details the 12 functions under test for Frama-C in
\cref{sec:xps}.  Second, \Cref{table-wp-ext} extends the results of
\cref{table-wp} and shows that WP is sensitive to all optimizations outside of
loop normalization.  Removing any of the 3 other optimizations affects
between 50\% and 66\% of the programs under study, rendering them non-provable.

\paragraph{Conclusion}
Extended experiments show that the 4 optimizations are complementary and
necessary to obtain the best of the analyzers.  Register unpacking has the
greatest impact on KLEE while high-level predicate recovery is necessary for the
integer abstractions used in Frama-C EVA. Removing any optimization outside of
loop normalization actually kills the effectiveness of Frama-C WP.

\clearpage
\newpage

\section{Additional experiments: size of produced code}
\label{sec:xp-ext-size}

The size of the generated code w.r.t the size of the assembly chunk is
indicative of what one can expect of the code produced by \approach/.
Since the size of the C code is also correlated with the quality of subsequent
analyses when produced by \approach/, this information becomes more notable.
Broadly speaking, the smaller the code, the easier it will be to analyze and the
better the analysis will be.

\medskip

\begin{table}[htbp]
  \caption{Ratio between C statements and assembly instructions w.r.t lifting
    strategies (all chunks, \x86/). }
  \centering
  \resizebox{\linewidth}{!}{
    \begin{tabular}{lcccccccc}
      &  \multicolumn{8}{c}{\sc Lifting} \\
      \light{\cmidrule(lr){2-9}}
       & {\sc Basic } & $O1$ & $O2$ & $O3$ & $O4$ & $\overline{O1}$ & $\overline{O2}$ & $\overline{O3}$ \\
      \light{\cmidrule(lr){2-2}\cmidrule(lr){3-3}\cmidrule(lr){4-4}\cmidrule(lr){5-5}\cmidrule(lr){6-6}\cmidrule(lr){7-7}\cmidrule(lr){8-8}\cmidrule(lr){9-9}}
      Min &  0.36 & 0.36 & 0.55 & 0.23 & 0.23 & 0.23 & 0.23 & 0.55 \\
      Average & 2.87 & 2.86 & 6.03 & 0.98 & 0.98 & 0.98 & 1.06 & 6.03 \\
      Max & 11.54 & 11.54 & 55.58 & 19.35 & 19.35 & 19.35 & 11.08 & 55.58 \\
    \end{tabular}}
  \label{tab:size-ratio}
\end{table}

\medskip

\Cref{tab:size-ratio} shows the ratio between the number of generated C
statements and the number of assembly instructions. As expected, two
optimizations significantly change the ratio. First, register unpacking
increases the number of generated statements as it splits single assignments
into many smaller assignments of independent values. On the other hand,
expression propagation greatly reduces the number of statements: on average,
this number decreases by a factor of 6 when it comes from register unpacking
and by a factor of 2.5 in other cases.  In the end, \approach/
roughly generates one C statement per assembly instruction for the chunks of the
Debian distribution.

\clearpage
\newpage

\section{\ffmpeg/ bug example}
\label{sec:example-ffmpeg}

\begin{figure}[htbp]
  \centering
  \begin{subfigure}[t]{\linewidth}
    \includegraphics{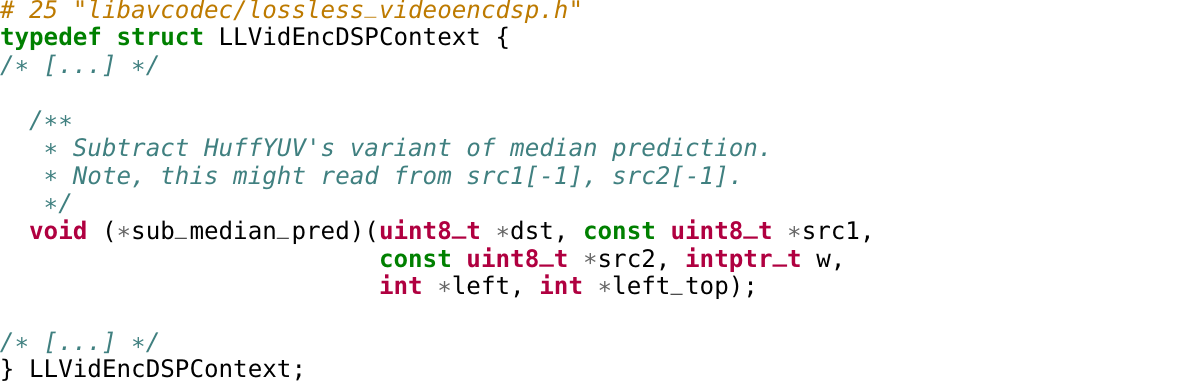}
    \caption{Function API}
    \label{fig:sub_median_pred_declaration}
  \end{subfigure}
  \begin{subfigure}[t]{\linewidth}
    \vspace*{.4cm}
    \hspace*{-20pt}\includegraphics{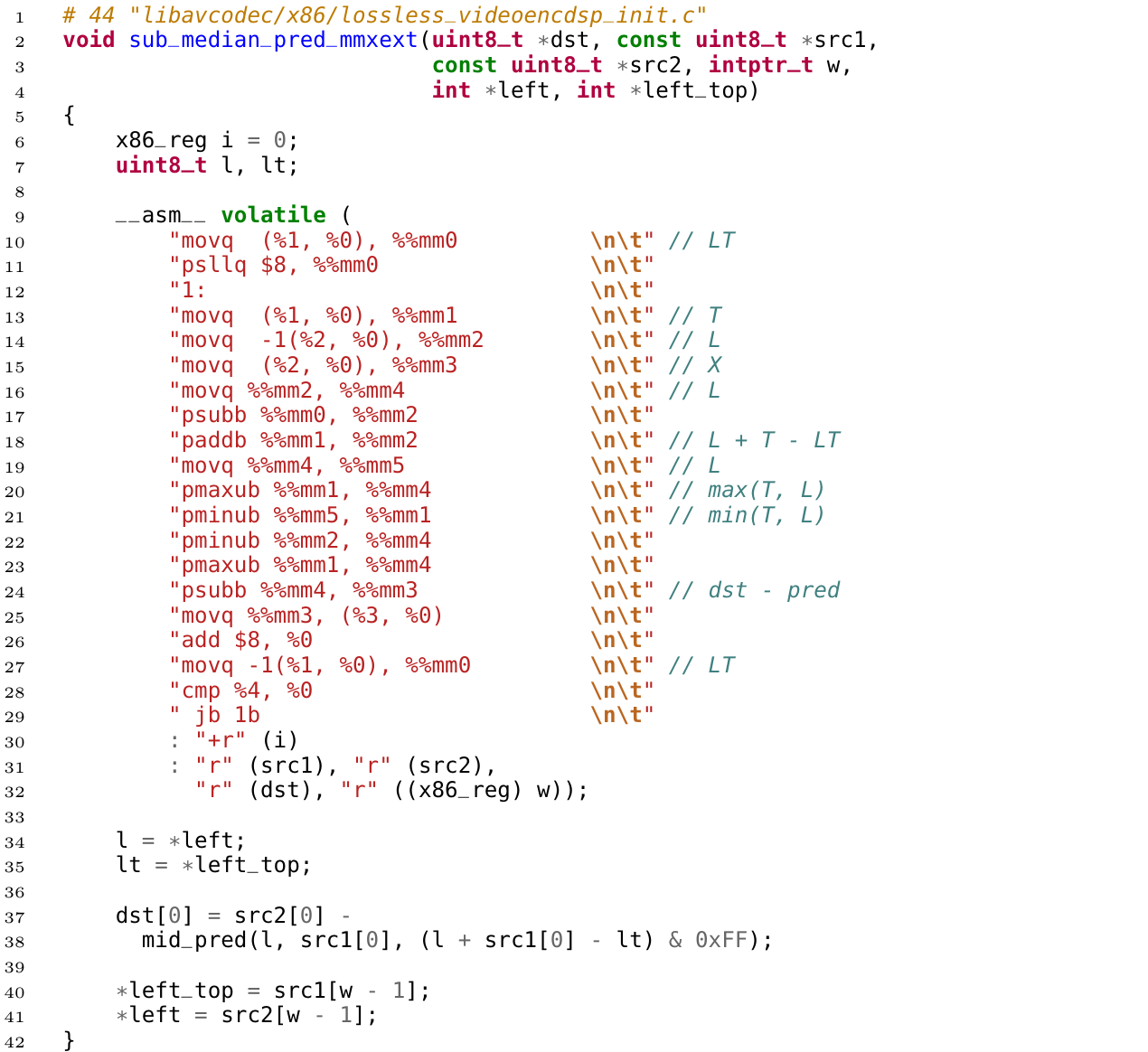}
    \caption{Original version}
    \label{fig:sub_median_pred_asm}
  \end{subfigure}
\end{figure}

\begin{figure*}[htbp]
  \ContinuedFloat
  \begin{subfigure}[t]{\linewidth}
    \hspace*{-20pt}\includegraphics{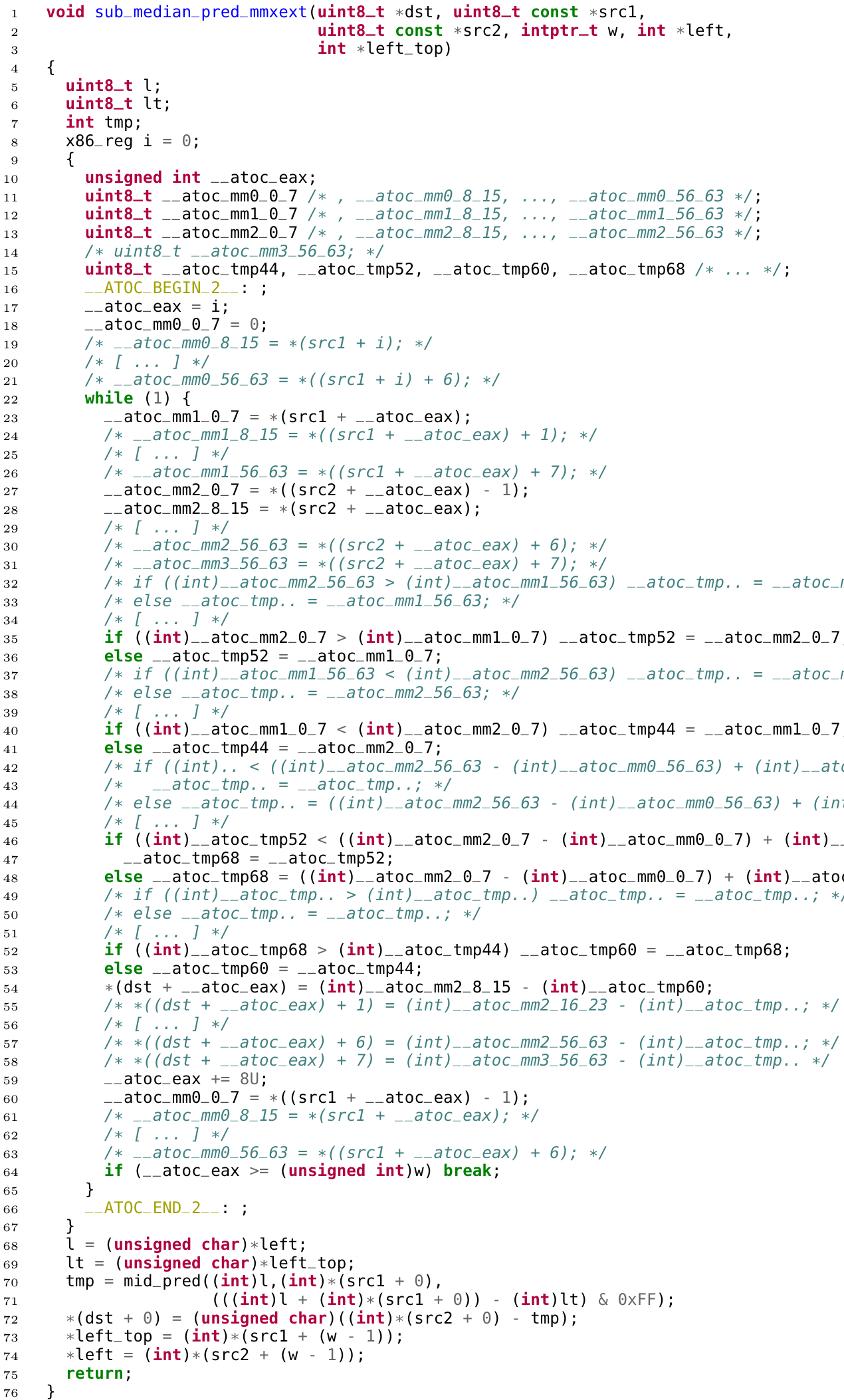}
    \caption{\approach/-generated version}
    \label{fig:sub_median_pred_tina}
  \end{subfigure}
  \caption{\ffmpeg/ function accessing \cinl{src2[-1]}}
  \label{fig:example-ffmpeg}
\end{figure*}

\Cref{sec:xp-findings} refers to a \ffmpeg/ function accessing
index \cinl{-1} of its input buffer. It is part of the lossless video
encryption provided by \ffmpeg/. More specifically, it is a field of
\cinl{struct LLVidEncDSPContext} (\cref{fig:sub_median_pred_declaration})
which is initialized dynamically, depending on the hardware capabilities.
Comments show that the developers know of this behavior.
\Cref{fig:sub_median_pred_asm,fig:sub_median_pred_tina} show the original
and generated version respectively of the function implementation using
\cinl{mmxext} extensions. The latter has been sliced for readability
reasons and the ellipsis stands for repetitive patterns.
(each MMX instruction is translated to 8 independent C statements)
The  \cinl{-1} access  occurs during the first iteration of the loop, when
\%0 (\cinl{eax}) is \cinl{0}, at line 14 (\cref{fig:sub_median_pred_asm})
in the assembly chunk. This bug is consequently found at line
27 in the lifted C version (\cref{fig:sub_median_pred_tina}).

\medskip

Since the situation is acknowledged in the code documentation, it is arguably
not a bug {\it sensu stricto}. However, the function is exported (not static),
does not contain defensive programming to avoid the bad behavior, and the
documentation is not directly on the function itself but in a record containing
a possible function pointer to this function.  Thus, we consider this a serious
programming flaw that could lead to issues down the line in several situations: code
maintenance and refactoring, code reuse (in other projects), compiler upgrades
(taking advantages of potential undefined behaviors to trigger more aggressive
code optimizations), etc.

\paragraph{Compliance issues}
Moreover, the chunk interface (\cref{fig:sub_median_pred_asm}, lines 30-32)
misses information about clobbering the \cinl{mm} registers 0 to 5 and accessing
memory from \cinl{src1}, \cinl{src2} and \cinl{dst}.  It is actually not an
issue {\it here}
as caller functions assumes (according to the Application Binary
Interface) that memory and \cinl{mm} registers are clobbered.  However, it is
not easy to know in advance how compilers will handle function inlining in terms of
memory write barriers and register clobbering. The missing information at the
interface of the chunk for clobbered entities could thus lead to serious issues when
compiling in a different environment. Similar concerns arise in the cases of
code reuse, especially in another project.

\clearpage

\section{Technical Focus on Simplification rules}
\label{sec-expression-propagation-ext}
%\begin{highlight}

We use a mixture of  standard and dedicated
simplification rules -- standard for typical integer-level properties and
dedicated for more low-level aspects.
In the rules below, we use the following notations
$|x|$ is the size of the expression $x$, $\diamond$ any binary operator, $C$ a condition ($|C| = 1$), $k$ is a constant.
$\vec{1}_{|x|}$ denotes a bitvector of size $|x|$ with all bits set to 1.
\begin{itemize}[]

\item standard ``lightweight'' term normalization in order to ease further other simplifications, including common subexpression elimination (a.k.a.~sharing),
unused variables elimination,  associativity-commutativity re-ordering.

\item constant propagation (modular arithmetic)

\item neutral elements:

  \begin{minipage}[t]{0.45\linewidth}
      \begin{align*}
    x + 0 &\hookrightarrow x \\
    x - 0 &\hookrightarrow x \\
    x \times 1 &\hookrightarrow x \\
    x \mathbin{\text{udiv}} 1 &\hookrightarrow x \\
    x \mathbin{\text{sdiv}} 1 &\hookrightarrow x \\
    x \mathbin{\text{urem}} 2^{|x|} &\hookrightarrow x
                                      \end{align*}
  \end{minipage}
  \begin{minipage}[t]{0.45\linewidth}
    \begin{align*}
      x \land \vec{1}_{|x|} &\hookrightarrow x \\
      x \vee 0 &\hookrightarrow x \\
      x \oplus 0 &\hookrightarrow x \\
      x \mathbin{\text{shl}} 0 &\hookrightarrow x \\
      x \mathbin{\text{shr}} 0 &\hookrightarrow x \\
      x \mathbin{\text{sar}} 0 &\hookrightarrow x  \\
    \end{align*}
  \end{minipage}

\item idempotence:

  \begin{minipage}[t]{0.45\linewidth}
    \begin{align*}
      x \land x &\hookrightarrow x \\
      x \vee x &\hookrightarrow x \\
      \text{uext}_{|x|}(x) &\hookrightarrow x \\
    \end{align*}
  \end{minipage}
  \begin{minipage}[t]{0.45\linewidth}
    \begin{align*}
      \text{sext}_{|x|}(x) &\hookrightarrow x \\
      \text{extract}_{0 .. |x| - 1}(x) &\hookrightarrow x
    \end{align*}
  \end{minipage}

\item absorbing(-like ) elements:

  \begin{minipage}[t]{0.45\linewidth}
    \begin{align*}
      x \times 0 &\hookrightarrow 0 \\
      x \land 0 &\hookrightarrow 0 \\
      x \vee 1 &\hookrightarrow 1 \\
    \end{align*}
  \end{minipage}
  \begin{minipage}[t]{0.45\linewidth}
    \begin{align*}
      x \mathbin{\text{urem}} 1 &\hookrightarrow 0 \\
      x \mathbin{\text{srem}} 1 &\hookrightarrow 0
    \end{align*}
  \end{minipage}

\item inverse elements:

  \begin{minipage}[t]{0.45\linewidth}
    \begin{align*}
      x - x &\hookrightarrow 0 \\
      x \mathbin{\text{udiv}} x &\hookrightarrow 1 \\
      x \mathbin{\text{shl}} k
                 &\overset{|x| \le k}{\hookrightarrow} 0  \\
    \end{align*}
  \end{minipage}
  \begin{minipage}[t]{0.45\linewidth}
    \begin{align*}
      x \oplus x &\hookrightarrow 0 \\
      x \mathbin{\text{sdiv}} x &\hookrightarrow 1 \\
      x \mathbin{\text{shr}} k
                 &\overset{|x| \le k}{\hookrightarrow} 0
    \end{align*}
  \end{minipage}

\item involutivity:

  \smallskip
  \hspace*{1cm} $\neg (\neg x) \hookrightarrow x $ \quad\quad\quad
  $- (- x) \hookrightarrow x $\hfill
  \smallskip

\item double shift simplifications:
  \begin{align*}
    (x \mathbin{\text{shl}} y) \mathbin{\text{shl}} z
    &\hookrightarrow x \mathbin{\text{shl}} (y + z) \\
    (x \mathbin{\text{shr}} y) \mathbin{\text{shr}} z
    &\hookrightarrow x \mathbin{\text{shr}} (y + z) \\
    (x \mathbin{\text{sar}} y) \mathbin{\text{sar}} z
    &\hookrightarrow x \mathbin{\text{sar}} (y + z)
  \end{align*}

  \item remainder / extension subsumption:
  \begin{align*}
         (x \mathbin{\text{urem}} k) \mathbin{\text{urem}} k'  &\overset{k \le k'}{\hookrightarrow} x \mathbin{\text{urem}} k \\
    (x \mathbin{\text{urem}} k) \mathbin{\text{urem}} k' &\overset{k > k'}{\hookrightarrow} x
                                           \mathbin{\text{urem}} k' \\
    \text{sext}_{k}(\text{uext}_{k'}(x)) &\overset{k' > |x|}{\hookrightarrow} \text{uext}_{k}(x) \\
    \text{uext}_{k}(\text{uext}_{k'}(x)) &\hookrightarrow \text{uext}_{k}(x)
  \end{align*}

\item condition simplifications

  \begin{minipage}[t]{0.45\linewidth}
    \begin{align*}
      C = 1 &\hookrightarrow C \\
      C \not = 0 &\hookrightarrow C \\
      C > 0 &\hookrightarrow C \\
      x = x &\hookrightarrow 1 \\
      x \not = x &\hookrightarrow 0 \\
    \end{align*}
  \end{minipage}
  \begin{minipage}[t]{0.45\linewidth}
    \begin{align*}
      x >_{u} x &\hookrightarrow 0 \\
      x <_{u} x &\hookrightarrow 0 \\
      x >_{s} x &\hookrightarrow 0 \\
      x <_{s} x &\hookrightarrow 0
    \end{align*}
  \end{minipage}

\item (extended) De Morgan simplifications

  \begin{minipage}[t]{0.45\linewidth}
    \begin{align*}
      \neg (C \land C') &\hookrightarrow \neg C \vee \neg C' \\
      \neg (C \vee C') &\hookrightarrow \neg C \land \neg C' \\
      x \oplus \vec{1}_{|x|} &\hookrightarrow \neg x  \\
      \neg (x = y) &\hookrightarrow x \not = y \\
      \neg (x \not = y) &\hookrightarrow x = y \\
      \neg (x <_{u} y) &\hookrightarrow y \le_{u} x \\
      \neg (x \le_{u} y) &\hookrightarrow y <_{u} x  \\
    \end{align*}
  \end{minipage}
  \begin{minipage}[t]{0.45\linewidth}
    \begin{align*}
      \neg (x >_{u} y) &\hookrightarrow y \ge_{u} x \\
      \neg (x \ge_{u} y) &\hookrightarrow y >_{u} x \\
      \neg (x <_{s} y) &\hookrightarrow y \le_{s} x \\
      \neg (x \le_{s} y) &\hookrightarrow y <_{s} x \\
      \neg (x >_{s} y) &\hookrightarrow y \ge_{s} x \\
      \neg (x \ge_{s} y) &\hookrightarrow y >_{s} x
    \end{align*}
  \end{minipage}

\item ternary expression simplification:
  \begin{align*}
        C \ ? \ false : true &\hookrightarrow \neg C \\
    \neg C \ ?\ x : y &\hookrightarrow C \ ?\ y : x \\
    C \ ? \ true : false &\hookrightarrow C \\
    C \ ?\ x : x &\hookrightarrow x \\
    x \diamond (C \ ? \ y : z) &\hookrightarrow
                                 C \ ? \ x \diamond y : x \diamond z \\
    (C \ ? \ w : x) \diamond (C \ ? \ y : z) &\hookrightarrow
                                               C \ ? \ w \diamond y : x \diamond z
  \end{align*}

\item split elements:
  \begin{align*}
    k = \text{concat}(x, y) & \hookrightarrow
                              (\text{extract}_{|y| .. |x| + |y| - 1}(k) = x) \\
                              &\quad \land
                              (\text{extract}_{0 .. |y| - 1}(k) = y) \\
    k \not = \text{concat}(x, y) & \hookrightarrow
                                (\text{extract}_{|y| .. |x| + |y| - 1}(k) \not =
                                   x) \\
                                   &\quad \vee
                                   (\text{extract}_{0 .. |y| - 1}(k) \not = y)
  \end{align*}

\item concatenation abstraction:
  \begin{align*}
    \text{uext}_{|x|+|y|}(x) \vee \text{concat}(y, 0_{|x|}) &\hookrightarrow \text{concat}(y, x) \\
    \text{uext}_{|x| + k}(x) \mathbin{\text{shl}} k &\hookrightarrow \text{concat}(x, 0_{k}) \\
    \text{concat}(0_{k}, x) &\hookrightarrow \text{uext}_{k + |x|}(x)
  \end{align*}

\item extraction simplification:
  \begin{align*}
        \text{extract}_{i .. j}(\text{extract}_{k .. l}(x))
    &\hookrightarrow \text{extract}_{i + k .. j + k}(x) \\
        \text{concat}(\text{extract}_{i .. j}(x),
    \text{extract}_{j + 1 .. k}(x)) &\hookrightarrow \text{extract}_{i .. k}(x) \\
            \text{extract}_{i .. j}(\text{uext}_{k}(x)) &\overset{|x| \le i}{\hookrightarrow} 0 \\
    \text{extract}_{0 .. |x| - 1}(x) &\hookrightarrow x \\
    \text{extract}_{i .. j}(\text{concat}(x, y))
     &\overset{j < |y|}{\hookrightarrow} \text{extract}_{i .. j}(y) \\
    \text{extract}_{i .. j}(\text{concat}(x, y))
 &\overset{|y| \le i}{\hookrightarrow}
                                            \text{extract}_{i - |y| .. j - |y|}(x) \\
    \text{extract}_{0 .. j}(\text{uext}_{k}(x))
    &\overset{|x| \le j}{\hookrightarrow} \text{uext}_{j}(x) \\
    \text{extract}_{i .. j}(\text{uext}_{k}(x))
      &\overset{j < |x|}{\hookrightarrow} \text{extract}_{i .. j}(x) \\
    \text{extract}_{0 .. j}(\text{sext}_{k}(x))
   &\overset{|x| \le j}{\hookrightarrow} \text{sext}_{j}(x) \\
    \text{extract}_{i .. j}(\text{sext}_{k}(x))
                                                        &\overset{j < |x|}{\hookrightarrow} \text{extract}_{i .. j}(x) \\
  \end{align*}

  \item two-complement arithmetic abstraction:
  \begin{align*}
    \neg x + 1 &\hookrightarrow - x \\
    (\text{uext}_{k}(x) \oplus 2^{|x| - 1}) - 2^{|x| - 1}
               &\hookrightarrow \text{sext}_{k}(x) \\
    \text{extract}_{|x| - 1}(x) &\hookrightarrow x <_{s} 0 \\
    \text{uext}_{n}(C) - 1 &\hookrightarrow C \ ? \ 0 : \vec{1}_{n} \\
    \text{sext}_{n}(C) &\hookrightarrow C \ ? \ \vec{1}_{n} : 0
  \end{align*}

\end{itemize}

\section{Loop Normalization example}
\label{sec:technical-focus-loop}

\Cref{fig:loop-norm1,fig:loop-norm2,fig:loop-norm3,fig:loop-norm4} illustrate
the 3 steps
on the motivating example. The \cinl{ecx} register stands as the loop counter,
the \cinl{edi} register is rebased, rescaled and unified with \cinl{ecx}.
Before merging \cinl{edi} with \cinl{ecx}, the following relation is recorded:
\cinl{edi} $\equiv$
\cinl{__tina_4} $+$~\cinl{4}~$\times$ \cinl{(__tina_3 - ecx)}.

\begin{figure}[htbp]
  \centering
  \begin{subfigure}[t]{\linewidth}
    \setlength{\tabcolsep}{1pt}
    \centering
    \begin{adjustbox}{width=3 \linewidth / 4,center}
      \begin{tikzpicture}[
          edge/.style={->, black, shorten >= 2pt, shorten <= 2pt},
          bb/.style={rectangle, draw, solid, rounded corners=1mm, minimum size=.7cm,
           inner sep=5pt},
        ]
        \coordinate (entry);
        \node[right = .1 of entry] {in};

        \node[bb, anchor = north] (a) at ($(entry) + (down:0.75)$)
             {\begin{tabular}{rcl}
                 \vout{ecx} & $\leftarrow$ & \vin{\_\_tina\_3} \\
                 \vout{edi} & $\leftarrow$ & \vin{\_\_tina\_4} \\
             \end{tabular}};

        \node[bb, anchor = north] (b) at ($(a.south) + (down:0.75)$)
             {\begin{tabular}{rcl}
                 \multicolumn{3}{c}{if \vin{ecx} = 0 then \textcolor{darkblue}{\sf break}} \\
             \end{tabular}};

        \node[bb, anchor = west, right = 0.5 of b] (c)
             {\begin{tabular}{rclcl}
                 \vout{@[}&\vin{edi}&\vout{]$_4$} & $\leftarrow$ & 0x00000000 \\
                 &\vout{edi}& & $\leftarrow$ & \vin{edi} $+$ 4 \\
                 &\vout{ecx}& & $\leftarrow$ & \vin{ecx} $-$ 1 \\
             \end{tabular}};

        \coordinate[below = .75 of b]  (exit);
        \node[right = .1 of exit] {out};

        \draw[edge, thick] (entry) -- (a.north);
        \draw[edge] (a.south) -- (b);
        \draw[edge] ([yshift=-5]b.east) -- ([yshift=-5]c.west);
        \draw[edge] ([yshift=5]c.west) -- ([yshift=5]b.east);
        \draw[edge, thick] (b) -- (exit);
      \end{tikzpicture}
    \end{adjustbox}
    \caption{Post $O3$}
    \label{fig:loop-norm0}
  \end{subfigure}
  \begin{subfigure}[t]{\linewidth}
    \setlength{\tabcolsep}{1pt}
    \centering
    \begin{adjustbox}{width=3 \linewidth / 4,center}
      \begin{tikzpicture}[
          edge/.style={->, black, shorten >= 2pt, shorten <= 2pt},
          bb/.style={rectangle, draw, solid, rounded corners=1mm, minimum size=.7cm,
           inner sep=5pt},
        ]
        \coordinate (entry);
        \node[right = .1 of entry] {in};

        \node[bb, anchor = north] (a) at ($(entry) + (down:0.75)$)
             {\begin{tabular}{rcl}
                 \vout{ecx} & $\leftarrow$ & \vin{\_\_tina\_3} \\
                 \vout{edi} & $\leftarrow$ & 0 \\
             \end{tabular}};

        \node[bb, anchor = north] (b) at ($(a.south) + (down:0.75)$)
             {\begin{tabular}{rcl}
                 \multicolumn{3}{c}{if \vin{ecx} = 0 then \textcolor{darkblue}{\sf break}} \\
             \end{tabular}};

        \node[bb, anchor = west, right = 0.5 of b] (c)
             {\begin{tabular}{rclcl}
                 \vout{@[}&\vin{\_\_tina\_4} $+$ \vin{edi}&\vout{]$_4$} & $\leftarrow$ & 0x00000000 \\
                 &\vout{edi}& & $\leftarrow$ & \vin{edi} $+$ 4 \\
                 &\vout{ecx}& & $\leftarrow$ & \vin{ecx} $-$ 1 \\
             \end{tabular}};

        \coordinate[below = .75 of b]  (exit);
        \node[right = .1 of exit] {out};

        \draw[edge, thick] (entry) -- (a.north);
        \draw[edge] (a.south) -- (b);
        \draw[edge] ([yshift=-5]b.east) -- ([yshift=-5]c.west);
        \draw[edge] ([yshift=5]c.west) -- ([yshift=5]b.east);
        \draw[edge, thick] (b) -- (exit);
      \end{tikzpicture}
    \end{adjustbox}
    \caption{Step 1: rebasing}
    \label{fig:loop-norm1}
  \end{subfigure}
  \begin{subfigure}[t]{\linewidth}
    \setlength{\tabcolsep}{1pt}
    \centering
    \begin{adjustbox}{width=3 \linewidth / 4,center}
      \begin{tikzpicture}[
          edge/.style={->, black, shorten >= 2pt, shorten <= 2pt},
          bb/.style={rectangle, draw, solid, rounded corners=1mm, minimum size=.7cm,
           inner sep=5pt},
        ]
        \coordinate (entry);
        \node[right = .1 of entry] {in};

        \node[bb, anchor = north] (a) at ($(entry) + (down:0.75)$)
             {\begin{tabular}{rcl}
                 \vout{ecx} & $\leftarrow$ & \vin{\_\_tina\_3} \\
                 \vout{edi} & $\leftarrow$ & 0 \\
             \end{tabular}};

        \node[bb, anchor = north] (b) at ($(a.south) + (down:0.75)$)
             {\begin{tabular}{rcl}
                 \multicolumn{3}{c}{if \vin{ecx} = 0 then \textcolor{darkblue}{\sf break}} \\
             \end{tabular}};

        \node[bb, anchor = west, right = 0.5 of b] (c)
             {\begin{tabular}{rclcl}
                 \vout{@[}&\vin{\_\_tina\_4} $+$ 4 $\times$ \vin{edi}&\vout{]$_4$} & $\leftarrow$ & 0x00000000 \\
                 &\vout{edi}& & $\leftarrow$ & \vin{edi} $+$ 1 \\
                 &\vout{ecx}& & $\leftarrow$ & \vin{ecx} $-$ 1 \\
             \end{tabular}};

        \coordinate[below = .75 of b]  (exit);
        \node[right = .1 of exit] {out};

        \draw[edge, thick] (entry) -- (a.north);
        \draw[edge] (a.south) -- (b);
        \draw[edge] ([yshift=-5]b.east) -- ([yshift=-5]c.west);
        \draw[edge] ([yshift=5]c.west) -- ([yshift=5]b.east);
        \draw[edge, thick] (b) -- (exit);
      \end{tikzpicture}
    \end{adjustbox}
    \caption{Step 2: rescaling}
    \label{fig:loop-norm2}
  \end{subfigure}
  \begin{subfigure}[t]{\linewidth}
    \setlength{\tabcolsep}{1pt}
    \centering
    \begin{adjustbox}{width=3 \linewidth / 4,center}
      \begin{tikzpicture}[
          edge/.style={->, black, shorten >= 2pt, shorten <= 2pt},
          bb/.style={rectangle, draw, solid, rounded corners=1mm, minimum size=.7cm,
           inner sep=5pt},
        ]
        \coordinate (entry);
        \node[right = .1 of entry] {in};

        \node[bb, anchor = north] (a) at ($(entry) + (down:0.75)$)
             {\begin{tabular}{rcl}
                 \vout{ecx} & $\leftarrow$ & \vin{\_\_tina\_3} \\
                 \vout{edi} & $\leftarrow$ & \vin{\_\_tina\_3} $-$ \vid{ecx} \\
             \end{tabular}};

        \node[bb, anchor = north] (b) at ($(a.south) + (down:0.75)$)
             {\begin{tabular}{rcl}
                 \multicolumn{3}{c}{if \vin{ecx} = 0 then \textcolor{darkblue}{\sf break}} \\
             \end{tabular}};

        \node[bb, anchor = west, right = 0.5 of b] (c)
             {\begin{tabular}{rclcl}
                 \vout{@[}&\vin{\_\_tina\_4} $+$ 4 $\times$ \vin{edi}&\vout{]$_4$} & $\leftarrow$ & 0x00000000 \\
                 &\vout{edi}& & $\leftarrow$ & \vin{\_\_tina\_3} $-$ \vin{ecx} $+$ 1 \\
                 &\vout{ecx}& & $\leftarrow$ & \vin{ecx} $-$ 1 \\
             \end{tabular}};

        \coordinate[below = .75 of b]  (exit);
        \node[right = .1 of exit] {out};

        \draw[edge, thick] (entry) -- (a.north);
        \draw[edge] (a.south) -- (b);
        \draw[edge] ([yshift=-5]b.east) -- ([yshift=-5]c.west);
        \draw[edge] ([yshift=5]c.west) -- ([yshift=5]b.east);
        \draw[edge, thick] (b) -- (exit);
      \end{tikzpicture}
    \end{adjustbox}
    \caption{Step 3: pre-merging}
    \label{fig:loop-norm3}
  \end{subfigure}
\begin{subfigure}[t]{\linewidth}
    \setlength{\tabcolsep}{1pt}
    \centering
    \begin{adjustbox}{width=3 \linewidth / 4,center}
      \begin{tikzpicture}[
          edge/.style={->, black, shorten >= 2pt, shorten <= 2pt},
          bb/.style={rectangle, draw, solid, rounded corners=1mm, minimum size=.7cm,
           inner sep=5pt},
        ]
        \coordinate (entry);
        \node[right = .1 of entry] {in};

        \node[bb, anchor = north] (a) at ($(entry) + (down:0.75)$)
             {\begin{tabular}{rcl}
                 \vout{ecx} & $\leftarrow$ & \vin{\_\_tina\_3} \\
             \end{tabular}};

        \node[bb, anchor = north] (b) at ($(a.south) + (down:0.75)$)
             {\begin{tabular}{rcl}
                 \multicolumn{3}{c}{if \vin{ecx} = 0 then \textcolor{darkblue}{\sf break}} \\
             \end{tabular}};

        \node[bb, anchor = west, right = 0.5 of b] (c)
             {\begin{tabular}{rclcl}
                 \vout{@[}&\vin{\_\_tina\_4} $+$ 4 $\times$ (\vin{\_\_tina\_3} $-$ \vin{ecx})&\vout{]$_4$} & $\leftarrow$ & 0x00000000 \\
                 &\vout{ecx}& & $\leftarrow$ & \vin{ecx} $-$ 1 \\
             \end{tabular}};

        \coordinate[below = .75 of b]  (exit);
        \node[right = .1 of exit] {out};

        \draw[edge, thick] (entry) -- (a.north);
        \draw[edge] (a.south) -- (b);
        \draw[edge] ([yshift=-5]b.east) -- ([yshift=-5]c.west);
        \draw[edge] ([yshift=5]c.west) -- ([yshift=5]b.east);
        \draw[edge, thick] (b) -- (exit);
      \end{tikzpicture}
    \end{adjustbox}
    \caption{Step 3: post-merging}
    \label{fig:loop-norm4}
  \end{subfigure}
  \caption{Loop normalization}
  \label{fig:loop-norm}
\end{figure}
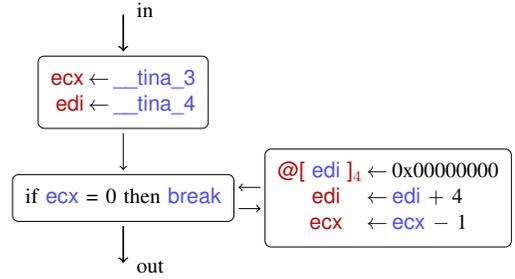
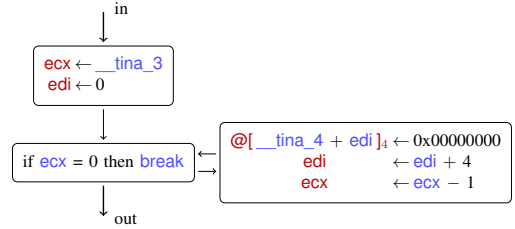
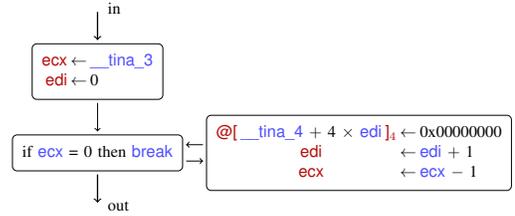
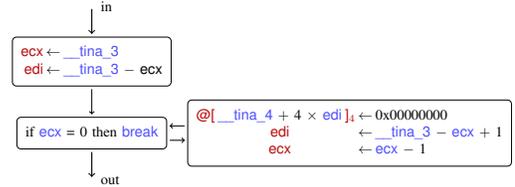
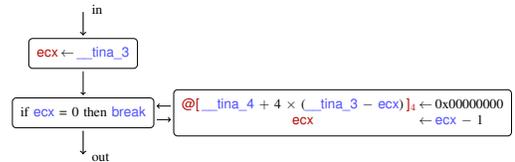

\section{Validation example}
\label{sec:technical-focus-validation}

  \Cref{fig:code-equivalence} shows the translation validation of the running
  example (\cref{fig:motivation_example}).  Codes before and after lifting have
  the same form (\ref{seq:isog}), as shown in~\cref{fig:eq:cfg}. Original blocks
  $B_0$, $B_1$ and $B_2$ have been paired respectively with their lifted
  counterparts $B'_0$, $B'_1$ and $B'_2$ for \ref{seq:block}.  Let us focus here
  on the equivalence check between $B_2$ and $B'_2$ (\cref{fig:eq:logic}).  The
  two blocks are obviously syntactically different, due to simplification and
  recompilation.  Compilation splits complex expressions using general registers
  as temporary variables while the ones from $B_2$ have been lifted to C
  variables with a close but different name. For instance \cinl{__tina_ecx} is
  equal to \cinl{ecx} whereas \cinl{eax} is used differently in $B_2$ and
  $B'_2$. Due to simplification passes, the two blocks no more have the same
  number of inputs or outputs, because lifting inferred some constraints and
  removed unused variables. Thus, \cinl{eax} and \cinl{df} are actually $0$ upon
  entering $B_2$. As pointer \cinl{edi} is incremented by $4$ while \cinl{ecx}
  is decremented by $1$, the two variables are linked by the linear relation
  \cinl{edi} $=$ \cinl{tina_4} $+$ \cinl{4} $\times$ (\cinl{tina_3} $-$
  \cinl{ecx}).  The formula is built to take into account these differences by
  adding logical assertions for each inferred constraint.  In the end, the
  equivalence query is then discharged by a SMT solver. An (expected) {\sf
    unsat} answer shows that there is no model such that the observable
  behaviors of the two basic blocks can differ.

\newcommand{\black}{\textcolor{black}}
\newcommand{\gray}[1]{{\color{gray} #1}}
\newcommand{\blue}{\textcolor{darkblue}}
\newcommand{\green}{\textcolor{green}}
\newcommand{\orange}{\textcolor{orange}}
\newcommand{\red}{\textcolor{darkred}}
\tikzstyle{bb}=
          [draw, rectangle, ultra thick, minimum width=100, minimum height=25]
\tikzstyle{edge}=[very thick, ->]

\newcommand{\fvid}[2]{\vid{#1}$_\text{#2}$}
\newcommand{\fvin}[1]{\fvid{\blue{#1}}{in}}
\newcommand{\fvout}[1]{\fvid{\red{#1}}{out}}

\begin{figure}[htbp]
  \begin{subfigure}[b]{.50\textwidth}
  {\centering
    \setlength{\tabcolsep}{1pt}
    \resizebox{\linewidth}{!}{
      \begin{tikzpicture}[
          edge/.style={->, black, shorten >= 2pt, shorten <= 2pt},
          bb/.style={rectangle, draw, solid, rounded corners=1mm, minimum size=.7cm,
           inner sep=5pt},
        ]
        \coordinate (entry);
        \node[right = .1 of entry] {in};
        \coordinate (entry') at ($(entry) + (right:6)$);
        \node[right = .1 of entry'] {in};

        \node[bb, anchor = north, below = .75 of entry] (a)
             {\begin{tabular}{rcl}
                 \vout{eax} & $\leftarrow$ & 0x00000000 \\
                 \vout{ecx} & $\leftarrow$ & \vin{tina\_3} \\
                 \vout{edi} & $\leftarrow$ & \vin{tina\_4} \\
                 \vout{df}  & $\leftarrow$ & 0 \\
             \end{tabular}};
        \node at ([xshift=-7]a.west) {$B_0$};

        \node[bb] (a') at ($(a) + (right:6)$)
             {\begin{tabular}{rcl}
                 {\sf eax} & $\leftarrow$ & \vin{tina\_3} \\
                 \vout{\_\_tina\_ecx} & $\leftarrow$ & {\sf eax} \\
             \end{tabular}};
        \node at ([xshift=-7]a'.west) {$B'_0$};

        \node[bb, anchor = north, below = .5 of a] (b)
             {\begin{tabular}{rcl}
                 \multicolumn{3}{c}{if \vin{ecx} = 0 then \blue{\sf break}} \\
             \end{tabular}};
        \node at ([xshift=-7]b.west) {$B_1$};

        \node[bb] (b') at ($(b) + (right:6)$)
             {\begin{tabular}{rcl}
                 \vid{zf} & $\leftarrow$ & \vin{\_\_tina\_ecx} = 0 \\
                 \multicolumn{3}{c}{if \vid{zf} then \blue{\sf break}}
             \end{tabular}};
        \node at ([xshift=-7]b'.west) {$B'_1$};

        \node[bb, anchor = north, below = 0.5 of b'] (c')
             {\begin{tabular}{rcl}
                 \vid{eax} & $\leftarrow$ & \vin{tina\_3} \\
                 \vid{eax} & $\leftarrow$ & \vid{eax} $-$ \vin{\_\_tina\_ecx} \\
                 \vid{edx} & $\leftarrow$ & 4 $\times$ \vid{eax} \\
                 \vid{eax} & $\leftarrow$ & \vin{tina\_4} \\
                 \vid{eax} & $\leftarrow$ & \vid{eax} $+$ \vid{edx} \\
                 \vout{@[}\vid{eax}\vout{]$_4$} & $\leftarrow$ & 0x00000000 \\
                 \vout{\_\_tina\_ecx} & $\leftarrow$ & \vin{\_\_tina\_ecx} $-$ 1 \\
             \end{tabular}};
        \node at ([xshift=-7]c'.west) {$B'_2$};

        \node[bb] (c) at ($(c') + (left:6)$)
             {\begin{tabular}{rclcl}
                 \vout{@[}\vin{edi}\vout{]$_4$} & $\leftarrow$ & \multicolumn{3}{l}{\vin{eax}} \\
                 \vout{edi} & $\leftarrow$ & \vin{df} & ? & \vin{edi} $-$ 4 \\
                            &              &          & : & \vin{edi} $+$ 4 \\
                 \vout{ecx} & $\leftarrow$ & \multicolumn{3}{l}{\vin{ecx} $-$ 1} \\
             \end{tabular}};
        \node at ([xshift=-7]c.west) {$B_2$};

        \coordinate[right = .75 of b]  (exit);
        \node[above = .1 of exit] {out};
        \coordinate[right = .75 of b'] (exit');
        \node[above = .1 of exit'] {out};

        \draw[edge, thick] (entry) -- (a);
        \draw[edge] (a) -- (b);
        \draw[edge] ([xshift=-5]b.south) -- ([xshift=-5]c.north);
        \draw[edge] ([xshift=5]c.north) -- ([xshift=5]b.south);
        \draw[edge, thick] (b) -- (exit);

        \draw[edge, thick] (entry') -- (a');
        \draw[edge] (a') -- (b');
        \draw[edge] ([xshift=-5]b'.south) -- ([xshift=-5]c'.north);
        \draw[edge] ([xshift=5]c'.north) -- ([xshift=5]b'.south);
        \draw[edge, thick] (b') -- (exit');
      \end{tikzpicture}
    }
  }
  \caption{Control flow graphs}
  \label{fig:eq:cfg}
  \end{subfigure}

  \begin{subfigure}[b]{\linewidth}
    \vspace*{1cm}
    \scriptsize
    \scriptsize\setlength{\tabcolsep}{3pt}
    \begin{tabular}{rcl}
      \multicolumn{3}{l}{; $B_2$ instructions sequence} \\
      \fvout{memory} & $=$ &
      store$_4$ \fvin{memory} \fvin{edi} \fvin{eax} \\
      \fvout{edi} & $=$ & if \fvin{df} then \fvin{edi} $-$ 4 else \fvin{edi} $+$ 4 \\
      \fvout{ecx} & $=$ & \fvin{ecx} $+$ 1 \\\\

      \multicolumn{3}{l}{; $B'_2$ instructions sequence} \\
      \fvid{eax'}{0} & $=$ & \fvin{tina\_3} \\
      \fvid{eax'}{1} & $=$ & \fvid{eax'}{0} $-$ \fvin{\_\_tina\_ecx} \\
      \fvid{edx'}{0} & $=$ & 4 $\times$ \fvid{eax'}{1} \\
      \fvid{eax'}{2} & $=$ & \fvin{tina\_4} \\
      \fvid{eax'}{3} & $=$ & \fvid{eax'}{2} $+$ \fvid{edx'}{0} \\
      \fvout{memory'} & $=$ &
      store$_4$ \fvin{memory} \fvid{eax'}{3} 0x00000000 \\
      \fvout{\_\_tina\_ecx} & $=$ &
      \fvin{\_\_tina\_ecx} $-$ 1 \\\\

      \multicolumn{3}{l}{; inputs restrictions} \\
      \fvin{eax} & $=$ & 0x00000000 \\
      \fvin{df}  & $=$ & \red{\sf false} \\
      \fvin{edi} & $=$ &
      \fvin{tina\_4} $+$ 4 $\times$ (\fvin{tina\_3} $-$ \fvin{ecx}) \\
      \fvin{ecx} & $=$ & \fvin{\_\_tina\_ecx} \\\\

      \multicolumn{3}{l}{; outputs restrictions} \\
      \fvid{ecx'}{out} & $=$ & \fvout{\_\_tina\_ecx} \\
      \fvid{edi'}{out} & $=$ &
      \fvin{tina\_4} $+$ 4 $\times$ (\fvin{tina\_3} $-$ \fvid{ecx'}{out}) \\\\

      \multicolumn{3}{l}{; outputs assertion} \\
      \multicolumn{3}{l}{\fvout{memory} $\not=$ \fvout{memory'}
        $\vee$ \fvout{ecx} $\not=$ \fvid{ecx'}{out}
        $\vee$ \fvout{edi} $\not=$ \fvid{edi'}{out}} \\
    \end{tabular}
    \caption{Logical formula for the equivalence between $B_2$ and $B'_2$}
    \label{fig:eq:logic}
  \end{subfigure}

  \caption{Basic block equivalence}
  \vspace*{-.5cm}
  \label{fig:code-equivalence}
\end{figure}

\end{appendices}

%%% Local Variables:
%%% mode: latex
%%% TeX-master: "main"
%%% End:

%  LocalWords:  Frama bitwise complementarity ACSL inlining bitvector
%  LocalWords:  subexpression associativity commutativity idempotence
%  LocalWords:  involutivity rebased rescaled rebasing rescaling SMT
%  LocalWords:  unsat

\end{long-paper}

\end{document}